\documentclass[]{article}
\usepackage{graphicx}
\usepackage{color}
\usepackage{listings}
\usepackage{fullpage}
\usepackage{amsmath}
\usepackage[utf8x]{inputenc}
\usepackage{import}
\usepackage{setspace}
\usepackage{authblk}
\usepackage{hyperref}
\usepackage{float}
\definecolor{lightgray}{gray}{0.5}
\usepackage{setspace}
\usepackage{color}
\usepackage[font={small,sf}, singlelinecheck=false]{caption}

\title{A compressed code for memory discrimination}

\author[1,2,3,*]{Dale Zhou}
\author[3]{Sharon M. Noh}
\author[4]{Nora C. Harhen}
\author[5]{Nidhi V. Banavar}
\author[6]{C. Brock Kirwan}
\author[1,2]{Michael A. Yassa}
\author[2,3]{Aaron M. Bornstein}

\affil[1]{University of California, Irvine, Neurobiology and Behavior, 519 Biological Sciences Quad, Irvine, 92697, United States}
\affil[2]{University of California, Irvine, Center for the Neurobiology of Learning and Memory, Qureshey Research Laboratory, Irvine, 92697, United States}
\affil[3]{University of California, Irvine, Department of Cognitive Sciences, Social Science Lab 334, Irvine, 92697, United States}
\affil[4]{New York University, Department of Psychology, 6 Washington Place, New York, NY 10003, United States}
\affil[5]{University of California, Berkeley, Department of Political Science, 210 Social Sciences Building, Berkeley, CA 94720, United States}
\affil[6]{University of Pennsylvania, MindCORE Neuroimaging Facility, 3401 Grays Ferry Ave, Philadelphia, 19146, United States}

\affil[*]{Corresponding author: \texttt{dale.zhou@uci.edu}}

\begin{document}

\maketitle

\clearpage
\newpage

\begin{abstract}

The ability to discriminate similar visual stimuli has been used as an important index of memory function. This ability is widely thought to be supported by expanding the dimensionality of relevant neural codes, such that neural representations for the similar stimuli are maximally distinct, or ``separated.'' An alternative hypothesis is that discrimination is supported by lossy compression of visual inputs, efficiently coding sensory information by discarding seemingly irrelevant details. A benefit of compression, relative to expansion, is that it allows the individual to efficiently retain fewer essential dimensions underlying stimulus variation---a process linked to higher-order visual processing---without hindering discrimination. Under the compression hypothesis, pattern separation is facilitated when more information from similar stimuli can be discarded, rather than preserving more information about distinct stimulus dimensions. We test the compression versus expansion hypotheses by predicting performance on the canonical mnemonic similarity task. First, we train neural networks to compress perceptual and semantic factors of stimuli, and measure lossiness of those representations using the mathematical framework underlying compression. Consistent with the compression hypothesis, and not the expansion hypothesis, we find that greater lossiness predicts the ease and performance of lure discrimination, particularly in later layers of convolutional neural networks shown to predict brain activity in the higher-order visual stream. We then empirically confirm these predictions across two sets of images, four behavioral datasets, and alternative metrics of lossiness. Finally, using task fMRI data, we identify signatures of lossy compression---neural dimensionality reduction and information loss---in the higher-order visual stream regions V4 and IT as well as hippocampal subregions dentate gyrus/CA3 and CA1 associated with lure discrimination performance. These results suggest lossy compression may support mnemonic discrimination behavior by discarding redundant and overlapping information.

\end{abstract}

\begin{quote}
\end{quote}

\section{Introduction}

\begin{figure}
\begin{center}
 \includegraphics[width=.9\columnwidth]{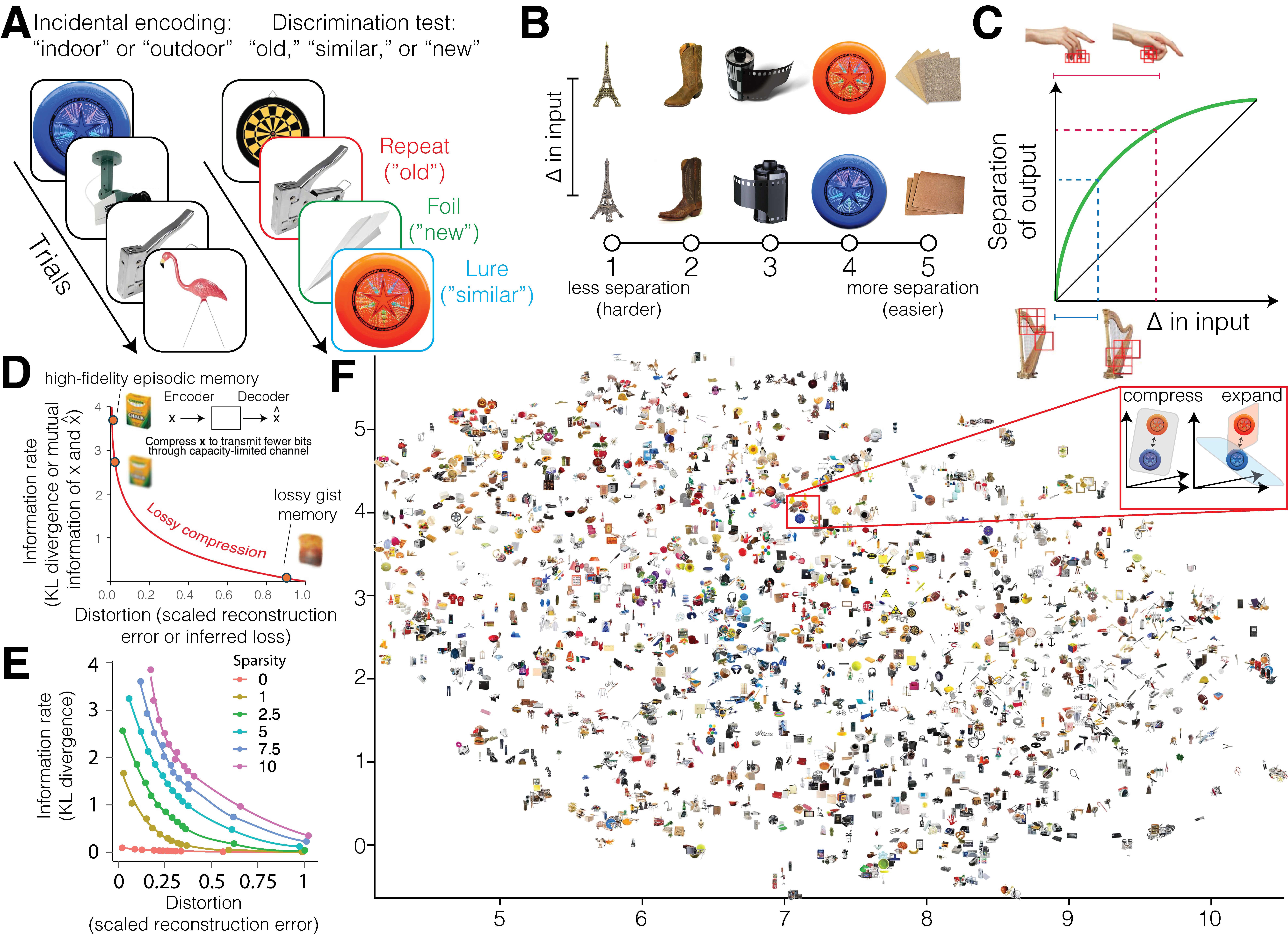}
\end{center}
\caption{\textbf{Task schematic and lossy compression function. }\textbf{(A)} Task stimuli during the encoding and test phases with corresponding trial type and correct response. \textbf{(B)} Ease of pattern separation between target and lure images. \textbf{(C)} Pattern separation function where differences in the input are encoded as distinct outputs. Distinguishing between extremely similar harps may require encoding of more information (blue-dashed lines) than distinguishing between more dissimilar hand gestures (red-dashed lines). Red bounding boxes indicate patches of maximal importance for discrimination by AlexNet. \textbf{(D)} Putative lossy compression computation supporting pattern separation by discarding similarities in inputs according to a rate-distortion function, the mathematical basis of compression. A high-fidelity memory might be associated with the semantic representation like: "a new yellow and green box of crayon chalk in white". A moderately compressed memory loses some detail: "a new yellow and green box of crayons." Finally, a very aggressively compressed memory loses essential details, resulting in false memory: "a used, crumpled yellow and red bag." \textbf{(E)} Key differences between the sparsity and compression hypotheses can be seen by increasing sparsity constraints on rate-distortion functions generated by $\beta$-variational autoencoders that are tasked with reconstructing each image. Data points are averaged across images per $\beta$. Sparsity makes lossy compression worse because it increases the information cost as well as the distortion by enforcing usage of a restricted set of encoding units. \textbf{(F)} Different neural networks lossily compress images into different feature representations. Here a UMAP representation space is visualized for images embedded by their semantic representation from a vision-text transformer. \textit{Inset}: How does lossily compressed dimensionality reduction (similar images represented in one plane) versus dimensionality expansion (similar images represented in more orthogonal planes) affect their memory discriminability?} 
\label{fig1}
\end{figure}

Many behaviors, from value-based decisions~\cite{noh2014multilevel, botvinick2015reinforcement, bornstein2023associative, pettine2023human} and associative learning~\cite{cayco-gajic_re-evaluating_2019} to perceptual inference and memory~\cite{hulbert2015neural, kim2017neural, khoudary2022precision, noh2023memory, bein2023predictions}, require recognizing whether the perception of a current situation is familiar or novel. This process is challenging because memory is constructive. From a percept with partial information, memory integrates prior experiences to fill in gaps but, in doing so, introduces distortions that can create incorrect impressions of prior experience~\cite{bartlett1995remembering, loftus1996memory}. Discriminating whether the current situation is novel compared to remembered experiences depends on pattern separation, the ability to distinguish between highly similar inputs with distinct responses \cite{yassa2011pattern, motley2012parametric}. Pattern separation occurs when brain regions, such as the hippocampus, transform similar inputs that would produce aligned activity patterns into output patterns that are more distinct \textbf{(Figure \ref{fig1}A-C)} ~\cite{bakker2008pattern, leutgeb2007pattern, bakker2008pattern, marr1991simple, o1994hippocampal, mcclelland1996considerations, o2001conjunctive, o2002hippocampal, suthana2015specific, dimsdale2018ca1, sakon2019neural, stark2019mnemonic}. Distinctness is often measured by the degree of linear independence where maximally distinct patterns are orthogonal. Although successful behavioral pattern separation is thought to be supported by reducing redundant overlap and keeping distinct details between inputs~\cite{treves1994computational, norman2010hippocampus, leal2018integrating}, it is not fully understood how computations support orthogonalization nor what input properties are orthogonalized~\cite{10.3389/fnbeh.2013.00096}.

The longstanding Marr-Albus hypothesis suggests two strategies to orthogonalize representations: expand the encoding ensemble and sparsen encoding activity~\cite{marr1969theory, albus1971theory}. Both expanding the number of encoding units (neurons) using divergent projections from a small to a large population (a ratio of about 1:5 in the hippocampus) and inhibiting the population activity to have few active neurons within a relevant timespan (around 5\% of neurons active) can decorrelate the statistical structure of inputs to separate across distinct ensembles of neurons~\cite{marr1969theory, albus1971theory, o1994hippocampal, leutgeb2007pattern, myers2009role, norman2010hippocampus, myers2011pattern, billings_network_2014, chavlis2017dendrites, cayco-gajic_re-evaluating_2019, guzman_how_2021}. However, expansion, sparsity, and decorrelation can have varying effects on pattern separation depending on the task and type of input, and it is difficult to disentangle their varying effects due to shared biological bases~\cite{chanales2017overlap, cayco-gajic_re-evaluating_2019}.

A less explored hypothesis for pattern separation is lossy compression, a computation that discards redundant information to produce efficient representations with a tolerable level of error~\cite{sims2018efficient, mack2020ventromedial, zhou2022efficient}. In contrast to the expansion strategy, compression suggests roles for reducing an encoding ensemble to create a physical or information bottleneck that encourages orthogonal representations of distinct features that dominate variation across inputs ~\cite{higgins2017betavae, higgins2021unsupervised}. In contrast to the sparsity strategy, compression benefits from different (sometimes lower) levels of sparsity due to better flexibility and expressivity with denser, mixed codes~\cite{rigotti2013importance}. Excessive sparsity can encumber code diversity while reducing sparsity can avoid oversensitive responses to inconsequential variations in the input~\cite{hasselmo2006role, gao2015simplicity, gao2017theory, stringer2019high, cohen2020separability, koolschijn2021memory}. 

The computational kinship between pattern separation and lossy compression can be made more apparent mathematically, distinguishing this account from alternative hypotheses such as sparsity \textbf{(Figure 1D-E)}. Pattern separation has been proposed to be the computation that decreases an arbitrary similarity metric, $S(X_1,X_2)$, of the degree of overlap between given inputs $X_1$ and $X_2$ ~\cite{10.3389/fnbeh.2013.00096}. Pattern separation occurs when $S(X_1,X_2)$ decreases across neural regions or neuronal populations $A$, $B$, and $C$, such that $X_1$ and $X_2$ are progressively decorrelated:
\begin{equation}
S_A(X_1, X_2) > S_B(X_1, X_2) > S_C(X_1, X_2).
\label{eq:decorrelation}
\end{equation}

\noindent Although $S$ is commonly conceived as a linear correlation, $S$ can also be defined as the mutual information $I$ to capture both linear and nonlinear dependencies between inputs. Under this definition of $S$, Equation \ref{eq:decorrelation} directly parallels the data processing inequality~\cite{10.5555/1146355}:

\begin{equation}
I_A(X_1; X_2) \geq I_B(X_1; X_2) \geq I_C(X_1; X_2),
\end{equation}

\noindent stating that physical processing from $A \rightarrow B \rightarrow C$ cannot create new information about the original source. The inequality points to a trade-off where information is either retained at some cost or lost for more lightweight but error-prone transmission, a core computational problem of memory. Memory needs to reconstruct arbitrary traces yet cannot preserve all of the information and structure of inputs~\cite{treves1994computational}. 

How many bits of information should be allocated to more precise high-fidelity memory versus saved for more approximate gist memory~\cite{nagy2020optimal, nagy2025interplay}? Lossy compression provides a framework to determine the optimal solution to this trade-off. Optimally, lossy compression is the joint minimization of (1) the \textit{rate} of information $R$ needed to encode an input $X_1$ as a compressed representation $X_2$ and (2) the amount of \textit{distortion} $D$ caused by information lost from compressing $X_1$ into $X_2$~\cite{shannon1959coding}: 
\begin{equation}
R(D) = \min I(X_1;X_2) \text{ subject to } d(X_1, X_2) \leq D. 
\end{equation}

\noindent 
Rate-distortion theory shows how lossy compression forces approximations such that $D>0$. We propose that reducing $S(X_1;X_2)$ for pattern separation involves minimizing existing redundancy, $I(X_1;X_2)$, which is marked by detectable increases in distortion $d_A(X_1, X_2) < d_B(X_1, X_2) < d_C(X_1, X_2)$.

Here, we test if lossy compression can explain performance in a behavioral task designed to tax pattern separation, the Mnemonic Similarity Task (MST)~\cite{stark2019mnemonic}. Participants incidentally encode information from a single exposure to a ``target'' image of an everyday object, then discriminate that memory from a similar yet distinct ``lure'' image and a novel ``foil'' image in a surprise discrimination test \textbf{(Figure \ref{fig1})}. Discrimination performance, measured by the proportion of correctly identified lures relative to foils, is thought to assess detail knowledge or specific recollection~\cite{norman2010hippocampus, yonelinas2010recollection, leal2014asymmetric}. We analyzed performance on 1,152 target-lure pairs across five previously published datasets~\cite{noh2021multi, nash2021pattern, noh2023memory, stark2023optimizing, banavar2024response}: a cross-sectional university sample ($n=208$), a longitudinal university sample that performed lure discrimination immediately and after 1 week ($n=78$), a cross-sectional sample of youths ($n=92$, ages 8 to 25, average of $15.80 \pm 5.12$ years), a cross-sectional lifespan aging sample ($n=297$, ages 18 to 86, average of $47.41 \pm 19.61$ years), and a cross-sectional sample who underwent fMRI scanning while performing the task ($n=48$, $22.9 \pm 3.6$ years old). We focus on two pairs of trials---corresponding target and lure trials, as well as first presentation and repeat trials. Using these data, we investigated how lossy compression contributes to the orthogonalization of target and lure images in support of pattern separation. Because visual information reaches the hippocampus after processing by visual and semantic cortical pathways~\cite{motley2012parametric}, we extracted various perceptual and/or semantic features by processing images through neural networks trained for pixel reconstruction (perceptual), image classification (perceptual and semantic), or image-to-text conversion (semantic)~\cite{higgins2017betavae, NIPS2012_c399862d, Simonyan2014VeryDC, lewis2015neural, Radford2021LearningTV}. These models have been predictive of the neural activity of temporal and visual cortices~\cite{yamins, benna2021place, wang_better_2023}, which may help separate inputs by transforming the dimensionality of representations~\cite{recanatesi2019dimensionality, cohen2020separability, farrell2022gradient}. The lossiness of compression is operationalized by several convergent approaches. In behavioral data, we use an information-theoretic algorithm that we modify to estimate lossiness from a cosine similarity metric of orthogonalization~\cite{sims2018efficient} and auto-associative networks that measure lossiness as item reconstruction errors~\cite{higgins2017betavae, alemi2017deep, nagy2020optimal, bates2020efficient}. In neural data, we use dimensionality reduction and the information rate (mutual information) of the evoked neural representations for targets and lures as operationalizations of lossy compression~\cite{mack2020ventromedial, zhou2022compression}. 

To accomplish pattern separation, the expansion hypothesis proposes that a sharpened representation of the total information reduces overlap, whereas the compression hypothesis posits that a blurred representation discards overlapping information. Across images, we test whether lossiness explains why pattern separation is sometimes more difficult, as previously defined by binning performance in an independent sample~\cite{lacy2011distinct} \textbf{(Figure \ref{fig1}B)}. Across individuals, we test if lossiness explains pattern separation performance, measured as the lure discrimination index: the proportion of correct rejection minus a response bias for ``similar.'' We hypothesize that pattern separation is more difficult and poorer when more bits of information are needed to preserve high-fidelity information about subtle differences in detail \textbf{(Fig \ref{fig1}C)}. Conversely, if the targets and lures are more dissimilar, then pattern separation is easier and better because lossy compression can aggressively discard more bits for greater efficiency. However, aggressive compression that discards more bits per unit of lossiness increases false alarms by blurring together gist-like memories which are more susceptible to noise (\textbf{Figure \ref{fig1}D)}. Finally, we examine whether stimulus-evoked responses in a putative hierarchy of neural regions reflects continually increasing or decreasing dimensionality, in line with the expansion or compression hypotheses. We find evidence in support of the lossy compression hypothesis, replicated across image datasets, image features, compression methods, and participant datasets. Lossier gist-like features capturing  higher-level perceptual features were more strongly related to pattern separation than low-level perceptual features, linking lossy compression with theories of object detection in the ventral visual stream by learning the most essential and invariant features in a lower-dimensional space~\cite{Rust12978, dicarlo_how_2012, kravitz2013ventral, cohen2020separability, CHUNG2021137}. Together, these findings support the idea that lossy compression, rather than expansion, supports pattern separation.

\section{Results}

\begin{figure}
\begin{center}
\includegraphics[width=1\columnwidth]{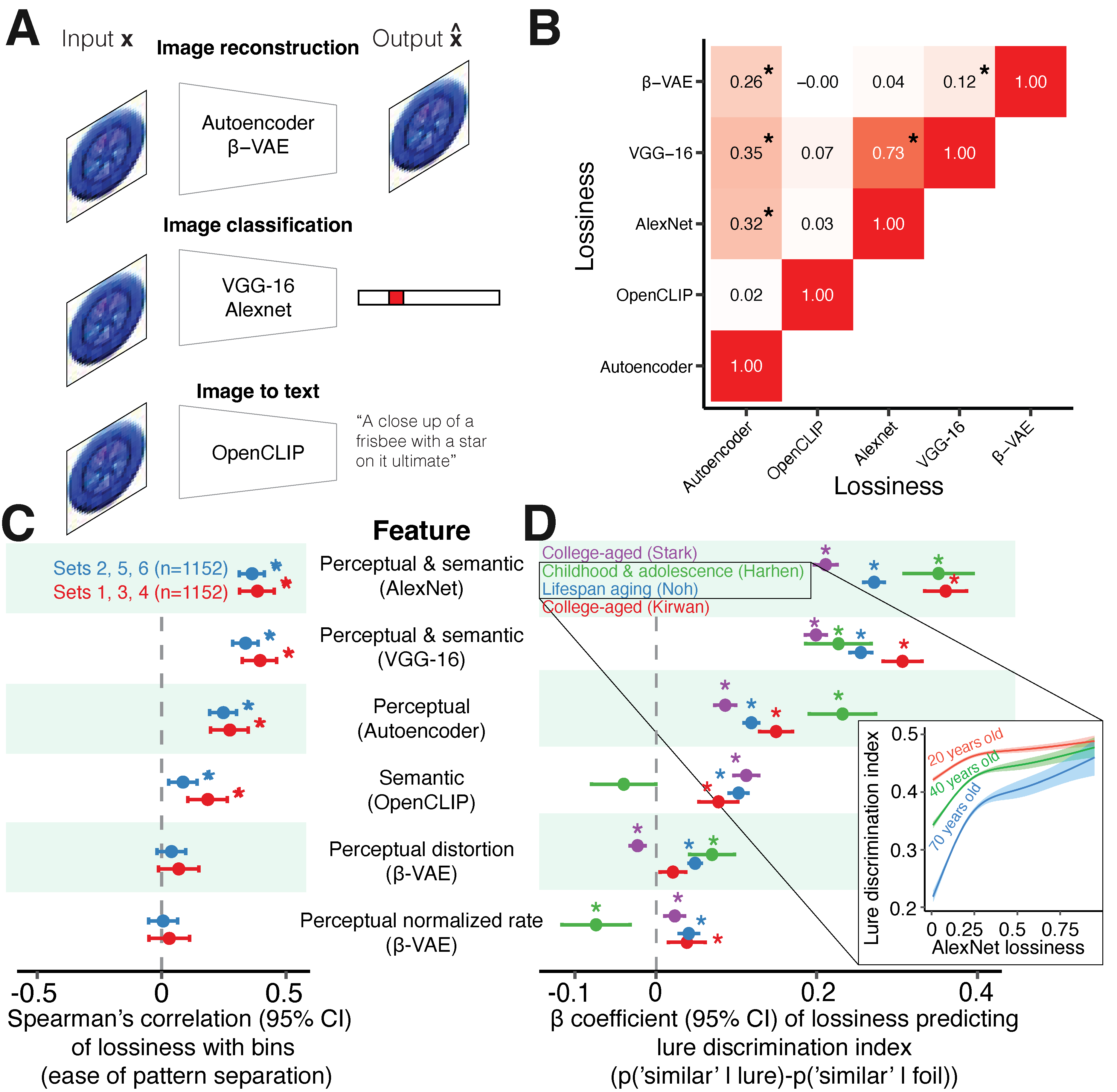}
\end{center}
\caption{\textbf{Lossiness of perceptual and semantic compression predicts pattern separation ease and performance. }\textbf{(A)} Neural networks labeled by their respective tasks. We applied each network to all images $x$ in order to obtain the sensory or semantic feature vectors $\hat{x}$ used to perform those tasks. \textbf{(B)} We measure the lossiness of compressing the $\hat{x}$ of a lure to the $\hat{x}$ of a target image by applying an information theoretic algorithm to simulated confusion matrices based on the cosine distance between $\hat{x}_{lure}$ and $\hat{x}_{target}$. The correlation table shows similarities and differences between the lossiness across models. \textbf{(C)} Consistent with our hypothesis, greater lossiness is correlated with easier pattern separation across two separate image sets. This effect was strongest for perceptual and sensory features. We did not observe an effect of lossiness using the $\beta$-VAE model. \textbf{(D)} Consistent with our hypotheses, greater lossiness is correlated with better pattern separation performance across datasets. The effect was greatest for perceptual and semantic features. The effect was less consistent for only semantic features and for the $\beta$-VAE model. Inset: nonparametric generalized additive modeling was used to flexibly model how lossiness and pattern separation interact with age. Individuals better pattern separate images with greater lossiness across the lifespan. The steeper logarithmic form for older adults ($F=732.1, p<0.001$, lossiness-by-age interactions $p<0.001$) suggests that lossiness may especially help reduce performance gaps between older compared to younger adults.} 
\label{fig3}
\end{figure}

\subsection*{Lossy compression relates to easier and better mnemonic pattern separation}

Can the lossiness of compressing inputs explain mnemonic pattern separation performance? We calculate the lossiness from the perceptual and semantic feature representations of the stimuli \textbf{(Figure \ref{fig3}A)}. Of particular interest are the pairs of stimuli evoking (1) a single-exposure memory of a target image during the study phase and (2) the subsequent exposure to a similar lure image during the test phase. Each neural network learns representations as a point in an internal model. Traversing the space of this internal model is a process of memory retrieval and reconstruction, where similar images are encoded more closely together according to the features learned by each neural network. A simple implementation of this traversal is a linear interpolation between points in the internal model. Using prior information theoretic methods on the generalization and discriminability of perceptual stimuli, we infer lossiness from a confusion matrix  constructed using the cosine distances between the points (targets, lures, and intermediate representations) of the retrieval and reconstruction process~\cite{sims2018efficient}. Lossiness values are similar for similar neural networks (e.g. AlexNet and VGG-16, Spearman's $\rho=0.73, p<0.001$) and uncorrelated across perceptual models and other generative models such as the OpenCLIP image-to-text transformer model ($\rho<0.07 , p>0.09$), suggesting good coverage of convergent and divergent representations of differing sensory and semantic features \textbf{(Figure \ref{fig3}B)}. 

Consistent with the compression hypotheses, we find that target and lure images that are easier to pattern separate tend to be those where compressing targets into lures has greater lossiness (all p-values Bonferroni corrected) \textbf{Figure \ref{fig3}C}). This effect reproduces across two image sets and all perceptual ($0.25<\rho<0.40$, $p<0.001$) and semantic models ($0.09<\rho<0.19$, $p<0.04$) that compress targets into lures, but not with the $\beta$-VAE models trained to reconstruct individual items ($\rho<0.07, p=1$). This suggests pattern separation is more related to the retrieval of compressed information about target and lure pairs, rather than efficient reconstruction of items in the pair themselves. We find further evidence consistent with the compression hypothesis, such that individuals tended to perform better when stimuli were more lossily compressed (linear regression $0.02<\beta<0.38$ Bonferroni-corrected $p<0.006$, \textbf{Figure \ref{fig3}D}). This effect largely replicates in 21 out of 24 tests across four datasets and the perceptual and semantic models. Next, we assess if age moderates the relationship between lossiness and lure discrimination performance. To this end, we used generalized additive models with penalized splines, a method which allows for statistically rigorous modeling of linear and nonlinear effects while minimizing over-fitting~\cite{wood2004stable}. Lure discrimination performance by older adults was more strongly related to the lossiness of compression than younger participants ($F=732.1, p<0.001$, lossiness-by-age interactions $p<0.001$), consistent with the notion that older adults can use semantic memory to compensate for degradations in episodic memory as their semantic knowledge increases \cite{park2009adaptive, naspi2023effects}. A small amount of lossiness can contribute significantly to improvement, but its benefits exhibit diminishing returns. In support of the compression hypothesis, lossiness predicts pattern separation ease and performance. Perceptual and semantic features were most related to the ease and performance of pattern separation.

\subsection*{Aggressive compression creates gist-like memories related to increased false alarms}

\begin{figure}
\begin{center}
 \includegraphics[width=1\columnwidth]{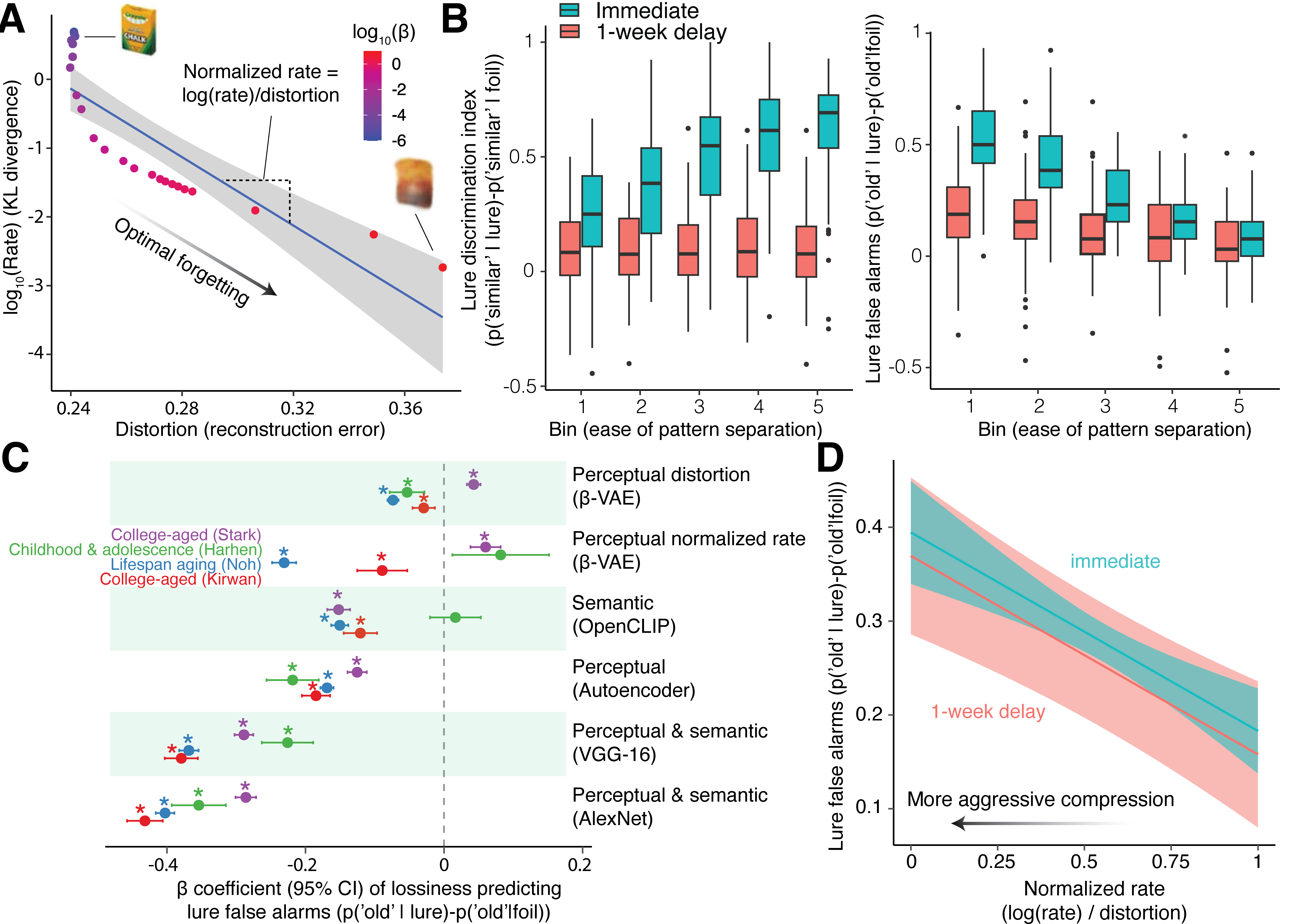}
\end{center}
\caption{\textbf{Gist-like memories due to aggressive compression explains increased false alarms. }\textbf{(A)} Each image has lossy compression applied such that the information rate and the reconstruction error between an input and output are optimized according to different rate-distortion trade-offs, $\beta$, using $\beta$-VAEs. Discarding information in a principled manner according to the information theory underlying lossy compression may characterizes a process of optimal forgetting by introducing adaptive distortions, but overly aggressive compression can lead to false memory due to gist-like representations (e.g., "a new yellow and green box of crayon chalk in white" versus "a used, crumpled yellow and red bag"). We test the benefits and drawbacks of distortion and the loss of information on lure false alarms. The optimal forgetting function can be quantified by calculating the slope of the rate-distortion function, which differs across images. We refer to this slope as the normalized rate because it is the amount of information discarded normalized by distortion for lossy compression. Steeper slopes indicate more aggressive compression that discards more information per unit of distortion, while shallower slopes indicate more conservative compression that preserves more information per unit of distortion. \textbf{(B)} \textit{Left:} Pattern separation performance degrades after 1 week. Pattern separation performance is better on easier stimuli in the immediate test but ease is not predictive of performance after 1 week. \textit{Right:} Lure false alarms occur more on harder stimuli in the immediate test but difficulty is not predictive of performance after 1 week. False alarms decreased after 1 week for harder stimuli. \textbf{(C)} Consistent with our hypothesis, greater lossiness across people is related to reduced lure false alarms, consistent with the idea that redundancy reduction decreases overlap across an individual's memories. \textbf{(D)} Consistent with our hypothesis, people viewing images with a lower normalized rate tended to have more lure false alarms in both the immediate and delayed tests. More aggressive compression discards more information for compact gist-like memory representations at the cost of a greater risk of false memory.}
\label{fig5}
\end{figure} 

Discarding information can enhance pattern separation by strategically targeting particular features of the image. Optimizing a neural network to retain essential features to remain distinguishable while compressing away non-critical correlations has been called a kind of ``optimal forgetting'' and adaptive distortion~\cite{nagy2020optimal, chanales2021adaptive}. While we already found some evidence for this hypothesis in pattern separation ease and performance, here we investigate whether the hypothesis also explains errors in performance. Errors in pattern separation are calculated by the lure false alarm rate, wherein individuals mistakenly confuse a lure for a previously studied image. We were also interested in how the different compressibility of unique images could explain how errors occur when there is memory interference introduced by a delayed test. Increasing lossiness corresponds to differing levels of information discarded according to rate-distortion curves unique to each image. We quantify this difference by calculating the normalized rate per image, or the slope of the rate-distortion function defined as the amount of information discarded per unit of distortion \textbf{(Figure \ref{fig5})A}. More negative, steeper slopes characterize forgetting that discards more information and resembles a more aggressive compression. More positive, shallower slopes characterize forgetting that retains more information and resembles a more conservative compression. While we test all datasets, we were particularly interested in the case of the longitudinal dataset where a substantial amount of interference and forgetting occurs after 1-week delayed test \textbf{(Figure \ref{fig5})B}. While the  trial difficulty no longer predicts the ease of pattern separation after 1 week, the normalized rate was associated with performance in both the immediate and delayed tests. Individuals tended to have worse pattern separation performance on images with more aggressive compression. Consistent with the compression hypothesis, greater lossiness correlated with lower lure false alarms \textbf{(Figure \ref{fig5})C}. This relationship largely replicated in 20 out of 24 tests across 4 datasets and the perceptual and semantic features tested. However, the effect of the normalized rate was less consistent, suggesting that lossiness is a more robust metric. 

In the longitudinal dataset, the ease of separating stimuli no longer predicts performance nor false alarms after a 1-week delayed test. Yet, a lower normalized rate was associated with more lure false alarms in both the immediate and delayed session ($\beta=-0.21, p<0.001$), indicating that aggressive compression can give rise to efficient but false gist-like memories. ``Optimally forgetting" information according to lossy compression appears to improve the overall discriminability of memories by reducing redundancies across representations; at the same time, aggressive compression can also generate false memories. 

\subsection*{Lossily compressed high-level perceptual and semantic representations associated with easier and better mnemonic pattern separation}

\begin{figure}
\begin{center}
\includegraphics[width=1\columnwidth]{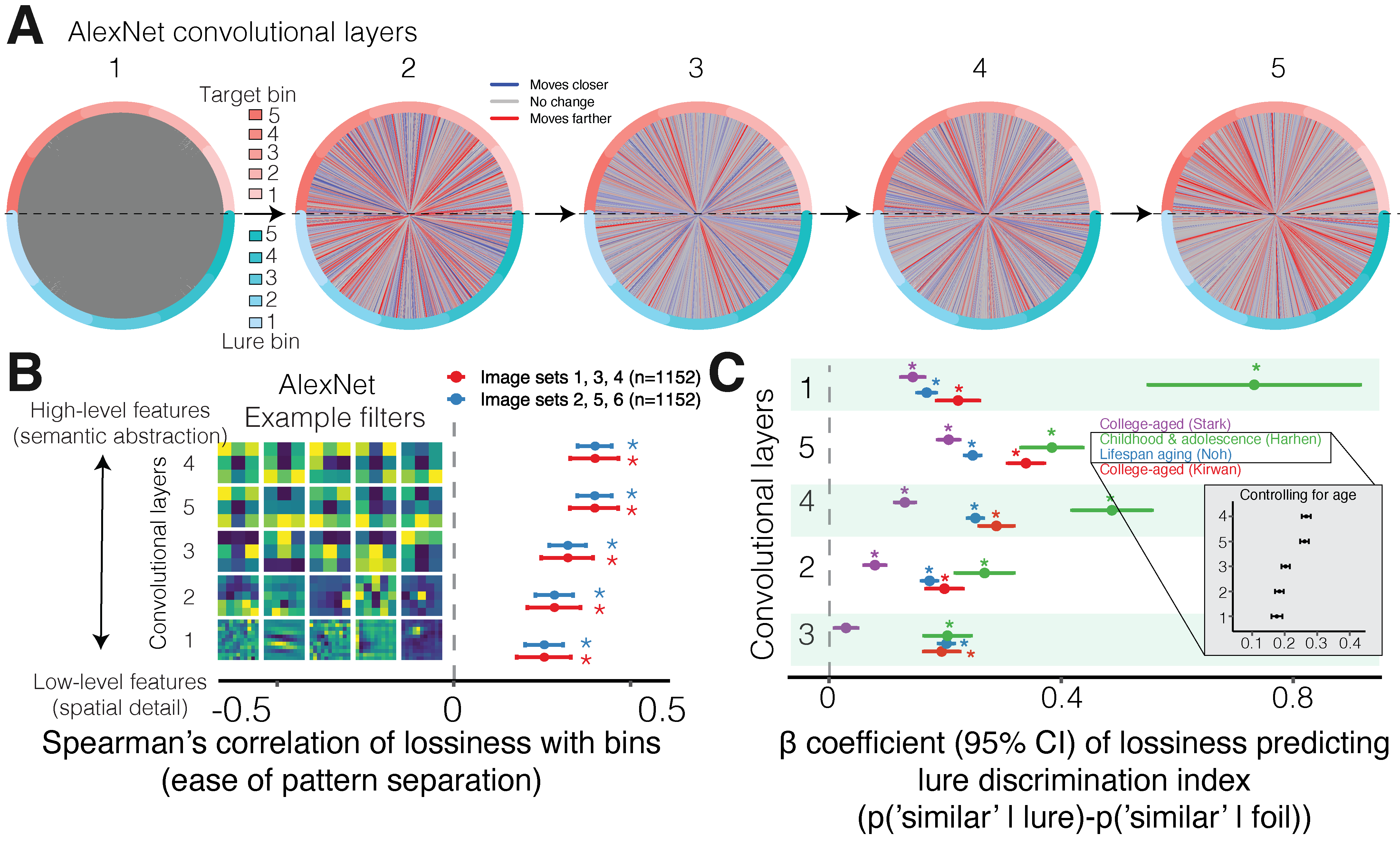}
\end{center}
\caption{\textbf{Lossiness of high-level semantic abstractions are more important than low-level spatial details for pattern separation ease and performance. }\textbf{(A)} What happens to the distance between target-lure pairs as they are processed through progressively deeper layers of AlexNet? Targets (pink) and lures (cyan) are shown as nodes, shaded by their ease of separation (bin). Edges indicate the relative change in Euclidean distance between target–lure pairs, based on dimensionality-reduced representations at each layer compared to the preceding one. Visually, earlier layers, known to process spatial details and orientation, appear to separate more difficult stimuli; later layers, known to process more semantically abstract information, appear to separate easier stimuli. \textbf{(B)} Left inset: AlexNet contains a hierarchy of convolutional layers that process different features of an image, including low-level perceptual features that represent edges, shapes, and textures (fine-grained filter); high-level perceptual features that represent conceptual and semantic abstractions for object categorization; and semantic features that represent verbal descriptions of images (coarse-grained filters). Right: Compressions of targets into lures features that incur greater lossiness tended to be better pattern separated. The lossiness of higher-level features characterizing semantic abstraction have a stronger effect on the ease of pattern separation than low-level features characterizing spatial detail. \textbf{(C)} Individual differences in pattern separation performance is most strongly predicted by the lowest-level layers for children and adolescents and highest-level layers of processing for older participants. Layers are visualized in descending order by average effect size. Inset: controlling for age results in the higher-level layers being more predictive of performance than lower-level layers.} 
\label{fig4}
\end{figure}

We next investigated which kinds of image features were most related to pattern separation ease and performance, comparing low-level features about spatial detail to high-level features about semantic abstraction. More compression of all perceptual features related to easier pattern separation ($0.26<\rho<0.40, p<0.001$). Higher-order semantic features (representations in deeper layers) had greater effects on the ease of pattern separation performance across image sets than lower-level features about spatial details (representations in shallower layers; $r=0.97, p<0.001$; \textbf{Figure \ref{fig4}A}). Moreover, we find a similar relationship when analyzing pattern separation performance ($0.08<\beta<0.83, p<0.001$ \textbf{Figure \ref{fig4}B}). Individuals tended to have better performance when either the deepest or most shallow features were more lossily compressed. 

Whether the deepest or most shallow features are more important may be driven by age. In the dataset of our youngest participants including children and adolescents, performance was more strongly related to low-level sensory details than higher-level features, as they may have not yet formed the richer semantic memories of older adults. To further test whether the order of importance of feature layers is explained by age, we conducted an analysis controlling for age as a covariate. After adjusting for age, the predictive strength of high-level features again surpassed that of low-level features \textbf{(\ref{fig4}B inset)}, indicating that age may moderate the observed shift from reliance on perceptual to semantic processing ($0.21<\beta<0.32, p<0.001$). This pattern of results suggests that lossily compressing high-level semantic features supports pattern separation, perhaps characterizing how the ventral visual stream discards inessential information to detect objects~\cite{Rust12978, dicarlo_how_2012, kravitz2013ventral, cohen2020separability, CHUNG2021137}. 

\subsection*{Neural signatures of lossy compression in higher-order visual stream and hippocampus associated with better pattern separation}

\begin{figure}
\begin{center}
 \includegraphics[width=1\columnwidth]{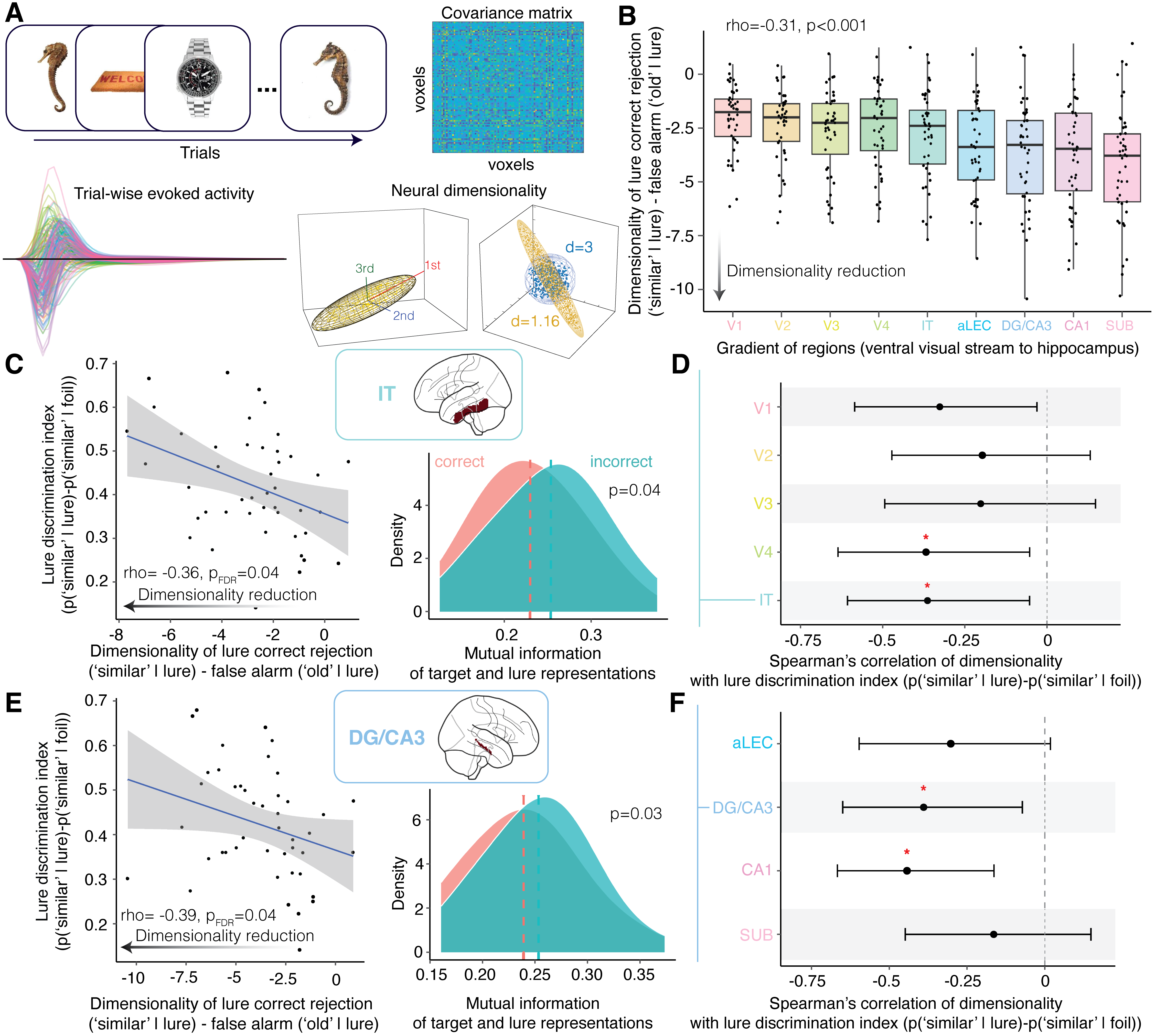}
\end{center}
\caption{\textbf{Signatures of lossy compression across the higher-order visual stream and hippocampus associated with pattern separation performance. }\textbf{(A)} \textit{Top-left}: Participants completed a continuous recognition paradigm, with no separate study and test phase for incidental encoding. \textit{Bottom-left}: For each trial, GLMsingle was used to estimate voxel-wise evoked activity magnitudes ($\beta$s) and an optimized hemodynamic response function from a library of canonical forms to repoduce trial-wise evoked activity time series ~\cite{prince2022improving}. Data visualized from one participant. \textit{Top-right}: The time series was used to calculate the voxel-wise covariance matrix. \textit{Bottom-right}: The magnitudes of principal components are the eigenvectors of the covariance matrix. The neural dimensionality (participation ratio) quantifies how many principal components effectively contribute to the variance captured along each direction (eigenvalues). When some voxels dominate the variance, there is low dimensionality. When voxels act more independently, there is higher dimensionality. \textbf{(B)} There is a gradient of progressively stronger dimensionality reduction from the ventral visual stream to the hippocampus. \textbf{(C-D)} Dimensionality reduction in higher-order visual regions in the ventral visual stream processing objects predicts pattern separation performance. In IT, correct trials had lower mutual information between target and lure reprsentations than incorrect trials. \textbf{(E-F)} Dimensionality reduction in hippocampal DG/CA3 and CA1 predict pattern separation performance. In DG/CA3, correct trials had lower mutual information between target and lure reprsentations than incorrect trials.}
\label{fig6}
\end{figure}

Lastly, in light of the computational and behavioral evidence associating lossy compression and pattern separation, we investigate neural signatures of lossy compression for separating evoked representations of target and lure stimuli \textbf{(Figure \ref{fig6})}. Two signatures were of interest: the dimensionality of the representations for target and lure stimuli pairs and the mutual information between the representations. While dimensionality expansion has been proposed to support pattern separation and object discrimination by increasing the neural separability of distinct clusters of information~\cite{cayco-gajic_re-evaluating_2019, CHUNG2021137}, the lossy compression framework predicts the opposite whereby dimensionality reduction helps to efficiently retain only a few of the most separable dimensions. Evidence supporting the compression hypothesis involves reduced dimensionality for correct trials (lure correct rejections) compared to incorrect trials (lure false alarms) that correlates with behavioral pattern separation performance. However, reduced dimensionality does not necessarily mean that there is less total information but rather that information is clustered around fewer dimensions. The compression hypothesis further predicts that the lower-dimensional representations exhibit a loss of information in correct trials compared to incorrect trials, a lower mutual information characteristic of lossiness. We tested 9 regions of interest, including V1, V2, V3, V4, IT, anterolateral entorhinal cortex, DG/CA3, CA1, and subiculum, as well as 2 reference regions that were not hypothesized to be directly involved in perceptual compression (primary somatosensory cortex and primary motor cortex). All comparisons are reported using FDR correction.

A circuit that performs pattern separation is proposed to exhibit a progressively stronger dissimilarity signature, such that input stimuli diverge more distinctly as population activity is measured across regions~\cite{10.3389/fnbeh.2013.00096}. Indeed, the dimensionality reduction for correct versus incorrect trials progressively strengthens from the ventral visual stream to the hippocampal circuit ($\rho=-0.31, p<0.001$; \textbf{Figure \ref{fig6}B}). Consistent with a compression hypothesis, we find that better pattern separation performance was associated with dimensionality reduction in higher-order visual stream regions V4 ($\rho=-0.37, p=0.043$; \textbf{Figure \ref{fig6}C-D}) and IT ($\rho=-0.36, p=0.043$; \textbf{Figure \ref{fig6}E-F}) as well as hippocampal DG/CA3 ($\rho=-0.39, p=0.043$) and CA1 ($\rho=-0.44, p=0.031$). We did not find any correlations between dimensionality expansion and pattern separation performance, nor did we find any associations involving the reference regions ($\rho>-0.29, p>0.09$). Bootstrap resampling with 10,000 iterations confirmed post-hoc that both M1 and S1 had significantly weaker correlations than the mean of other regions related to pattern separation: M1 vs. others = -0.10 [95\% CI: 0.08, 0.13], $p < 0.001$; S1 vs. others = -0.17 [95\% CI: 0.15, 0.21], $p < 0.001$). Together, the lossy compression of semantic abstractions and combinations of a few discriminable perceptual features in the ventral visual stream may support pattern separation in the hippocampus. 

\section*{Discussion}

In summary, we found behavioral, computational, and neural evidence supporting a lossy compression account of memory discrimination. The lossiness of compressing the semantic and perceptual input features explained the ease, performance, and errors in pattern separation. The relationship between lossiness and pattern separation was strongest when semantic abstractions were distilled from perceptual details in networks trained to perform image classification. This result is consistent with how the ventral visual stream first processes low-level spatial detail then high-level semantic abstractions to perform object detection~\cite{dicarlo_how_2012, kravitz2013ventral, yamins}. Moreover, layer-wise dimensionality reduction---a form of lossy compression---improves the separability of inputs in intermediate layers of neural network models associated with the later regions in the ventral visual stream~\cite{cohen2020separability, yamins, benna2021place, wang_better_2023}. Consistent with this computational result, we found that the dimensionality reduction of neural activity was associated with better pattern separation in high-level V4 and IT, but not low-level, visual stream regions for intermediate processing of abstract, semantic qualities of objects that are then later input into and further reduced within hippocampal DG/CA3 and CA1 to support pattern separation. The importance of high-level versus low-level information may be related to age. Young participants' performance was uniquely related to the lossiness of the lowest-level features, a relationship that was moderated with age. Pattern separation in children may be supported by the neurodevelopment of high-resolution memories for perceptual details~\cite{ ngo2018ontogeny, ngo2019building, canada2019s}, while loss of precision in mnemonic representations is related to episodic memory decline during aging~\cite{korkki2020healthy}. Indeed, performance in our youngest sample including children most related to the compression of perceptual details. Our results suggest that perceptual inputs which are more amenable to lossy compression can produce representations with strategically placed imprecision that still retain some essential dimensions that support memory. Consistent with this idea, older adults' performance was closer to that of younger adults when the stimuli had lossier compression of perceptual and semantic features. These results are consistent with the notion that the development of semantic knowledge may help organize stimuli in memory, providing a scaffold for perceptual memory encoding~\cite{rosch1975cognitive, luo2007age, dubova2021influences, kramer2023features, cohen2025pattern, nemeczdifferent}. The lossy compression framework explains why perceptual details should be lost faster than higher-level semantic constructs for memory discrimination.

Due to optimizing how overlapping information is discarded by separating input features~\cite{ganguli2012compressed, higgins2017betavae, higgins2021unsupervised}, lossy compression can be viewed as a process of ``optimal forgetting'' for different computational goals~\cite{ alemi2017deep, saxe2019information, nagy2020optimal}. The distortion caused by discarding information is not simply noise but is an adaptive distortion or error signal that can serve useful memory computations~\cite{schacter1999seven, wimber2015retrieval, sims2018efficient, wang2019more, lynn2020abstract, bates2020efficient, nagy2020optimal, chanales2021adaptive, lin2024images, schacter20241373, nagy2025interplay}. A variety of memory and learning phenomena can be understood under a shared framework by viewing memories as lossy compressions by the hippocampus ~\cite{treves1994computational, chanales2017overlap, wanjia2021abrupt, benna2021place, spens2024generative}, prefrontal, and perceptual processing streams which disentangle or repulse inputs to create the distances that support pattern separation~\cite{Rust12978, chanales2017overlap, chanales2021adaptive, mack2020ventromedial, 10.3389/fnbeh.2013.00096, motley2012parametric, reagh2014object, pidgeon2016cortical, wais2017retrieval, reagh2018functional, zhao2021adaptive, johnson2021rodent, CHUNG2021137, nash2021pattern, amer2023extra, wanjia2025repulsion}.  Indeed, the lossiness of compressing perceptually similar stimuli explains when those stimuli are more perceptually discriminable~\cite{sims2018efficient}, as the sensory system can afford more aggressive compression without needing to invest additional resources to retain finer details with high precision. 

These results can be understood more broadly in the context of the efficient coding principle. Efficient coding is when the brain maximizes the amount of information transmitted in an economical form by reducing redundancy~\cite{barlow1961possible, attneave1954some}. This coding strategy maximizes efficiency by allocating limited resources to where they are most needed for the task at hand~\cite{barlow1961possible}. For example, information is believed to be compressed even before the primary visual cortex, where $10^9$ bits/second in the photoreceptors are compressed to $10^7$ bits/second across retinal ganglion cells subject to the constraints of the capacity-limited optic nerve~\cite{li2014understanding}, and in a manner that is metabolically efficient~\cite{koch2006much, pitkow2012decorrelation, harris2015energy, zhou2022efficient}. Arriving into the visual cortex, signals processed by neural network models of ventral visual stream computation suggest progressive dimensionality reduction of inputs~\cite{cohen2020separability}. However the efficient coding principle should not be read as always compressing or reducing, but adaptively switching between compression and expansion modes depends on the complexity and dimensionality of the task~\cite{gao2017theory, mack2020ventromedial, owen2024high}. For instance, in our task we tested rapid, single-shot encoding from only one exposure to pairs of stimuli, whereas other tasks tax multiple repetitions and sessions across multi-dimensional sets of stimuli~\cite{tang2019effective}.  Indeed, compression and expansion computations both support learning because both help effectively orthogonalize representations dependent on the intrinsic dimensionality of the task and its inputs~\cite{alemi2017deep, saxe2019information, kolchinsky2019nonlinear, recanatesi2019dimensionality, farrell2022gradient, guzman_how_2021, zhou2022compression, zhou2022efficient, dubova2023excess, ito2023multitask, nigam2024predictions}. Hence, switching between differing levels of expansion and reduction may complement switching between different levels of inhibition and sparsity in support of adaptive memory function~\cite{hasselmo2006role, whittington2022disentanglement}.

In the context of memory and learning, compression and expansion processes can transform the dimensionality of the memory representations to support their separability or generalization~\cite{kerren2025exploring}. Such neural processes in the IT and perirhinal cortex may separate representations by category, novelty, and familiarity~\cite{ranganath2003neural, jaegle2019visual, nigam2024predictions}. This separability hinges on the transformation of representational space creating separation due to the change in dimensionality defining the location of representational points in the space~\cite{CHUNG2021137, cayco-gajic_re-evaluating_2019}. Novelty detection is thought to involve ventral tegmental and hippocampal loops that compare predictions generated by the memory traces in hippocampal regions with the new inputs from the cortex~\cite{lisman2005hippocampal, hasselmo1997free}. A future experiment could build upon our methods by using supervised or self-supervised redundancy reduction methods based on an information bottleneck approach to compression~\cite{kolchinsky2019nonlinear, zbontar2021barlow}. Selective attention to task-relevant factors shape perceptual feature representations in a similar manner as top-down modulation by hippocampal representations for memory and learning~\cite{carr2013top, aly2016attention, mack2020ventromedial, nemeczdifferent, son2024brief, mack2024discrimination}. In addition to reconstructed retrieval and top-down signals generated by the hippocampus and medial temporal region, regions earlier in the processing hierarchy in the visual and orbitofrontal cortex can support single-exposure perceptual memories that may be lossily compressed to support pattern separation~\cite{bar2006top, hindy2016linking, kim2020functions}. Eye-tracking, information bottleneck, and neural recordings could help elucidate how early compressive computations can support the pathway of pattern separation processes. The efficient coding hypothesis predicts sampling and storage of highly relevant features of information. For example, future work could investigate relationships between optimal features to remember and the features contributing to the memorability of objects~\cite{isola2011understanding, khosla2012memorability, rust2020understanding, bylinskii2021memorability, kramer2023features, revsine2025memorability}. Tantalizingly these approaches can provide a precise and general framework that explains how such salient features are extracted and later used if they are \textit{relevant}, where relevance is defined in the information theoretic terms of how much information can efficiently retained and discarded according to task demands~\cite{zbontar2021barlow, murphy2022interpretability, fang2025humans}. This relevance computation can be used to unify computational and neural processes linking memory,  control~\cite{kahn2004functional, noh2014multilevel, kuhl2013dissociable, richter2016distinct, favila2018parietal, nash2021pattern, yang2024pattern}, and category learning~\cite{nosofsky2000exemplar, sigala2002visual, folstein2013category, folstein2015category}. 

While compression and expansion have complementary functions, compression may prove more useful when stimuli contain redundancy (already have intermediate or heavy overlap) and are intrinsically low-dimensional~\cite{chanales2017overlap, cohen2020separability, CHUNG2021137}. In our behavioral pattern separation task, key stimulus dimensions include orientation, color, brightness, detail, number, scale, and/or shape factors~\cite{nemeczdifferent}. Compressed representations may support pattern separation in the dentate gyrus of the hippocampus for rapid, one-shot learning~\cite{benna2021place, o2001conjunctive, klippenstein2020neural}, whereas more prolonged processing is supported by activity in the ventral visual stream, determining the parts of the input that are invariant to changes in these dimensions, and other regions requiring larger changes in inputs across longer timespans~\cite{mcclelland1996considerations, norman2003modeling, rotshtein2005morphing, Rust12978, dicarlo_how_2012, motley2012parametric, kravitz2013ventral, pidgeon2016cortical, tang2019effective, nash2021pattern}. Prior work has already explored how regions contribute to pattern separation beyond the hippocampus according to complementary learning systems theory~\cite{motley2012parametric, paleja2014two, suthana2015specific, pidgeon2016cortical, klippenstein2020neural, nash2021pattern, gattas2023theta, yang2024pattern}. What the compression framework provides is a cohesive mathematical account for when differing levels of expansion/reduction and sparsity can directly and indirectly support pattern separation and related processes of general recognition, detailed recollection, and perceptual discrimination across the brain~\cite{10.3389/fnbeh.2013.00096, reagh2014object, reagh2018functional, davidson2019older, amer2023extra, johnson2021rodent, nash2021pattern}.

Our results can also be interpreted from a retrieval and pattern completion account, as memory discrimination tasks are not process pure~\cite{yassa2011pattern}. Pattern completion is the reactivation of a stored memory trace in response to a partial cue~\cite{bakker2008pattern}. Closely related to efficient coding and our compression methods is the analysis-by-synthesis theory, which posits that perception involves inferring a compact set of latent causes that can reconstruct the input~\cite{barrow1978computer, yuille2006vision, olshausen2014perception}. Such inference parallels memory reconstruction, where the learned latent structure enables efficient recall. The reconstruction error or distortion from this process characterize the (im)precision of the memory representation which can serve as an uncertainty or confidence signal for learning and recollection~\cite{harlow2016distinguishing}, further shaping both perception and memory~\cite{lin2024images, jang2024improved}. 

Computationally, retrieval and completion perspectives offer a complementary view to compressed encoding by accounting for how recollection processes supports memory discrimination and novelty detection via a recall-to-reject process~\cite{rotello2000associative, gallo2006two, norman2010hippocampus, yonelinas2010recollection, wahlheimrole, wahlheimmnemonic, DiRisio2025.09.22.677855}. Neurally, later regions in the ventral visual and occipital temporal stream as well as hippocampal CA3 and CA1 are regions involved in pattern completion and retrieval processes~\cite{wheeler2000memory, kuhl2011fidelity}, overlapping with the regions we identified. Our framework, like complementary systems models~\cite{marr1991simple, mcnaughton1995there}, involve both encoding and retrieval processes: inputs are compressed at encoding into a latent space, where latent variables are the ``causes'' of pixels in the stimuli. Traversing a distance in the latent space is akin to using the latent variables as a partial cue to reconstruct and recollect an associated stimulus (target-lure pair). What our framework provides is a normative model to determine the nature of what semantic or perceptual details should be encoded, retrieved, or discarded and at what level of precision or distortion given particular goals of a task. Future work could investigate these processes by quantifying the information cost of reconstruction given the partially distorted compression as a cue compared to attentional templates or memory schemas that represent goals, interacting with the memory and control processes of frontoparietal and association regions~\cite{kahn2004functional, kuhl2013dissociable, richter2016distinct, favila2018parietal, nash2021pattern, yang2024pattern, bein2025schemas}. 

This work has several limitations, some of which motivate future research. First, lossiness was inferred from an information theoretic method developed to fit empirical data from perceptual identification experiments, such as on a range of tones, colors, line lengths, and shapes~\cite{sims2018efficient}. To adapt this method to our pattern separation task, we simulated confusion matrices on a range of memory representations according to a putative retrieval process that linearly interpolates over cosine distances between internal memory representations. A consequence is that, in contrast to prior work showing that lossy compression provides an alternative to the theories dependent on psychological distance representations such as multidimensional scaling, our inferences of lossiness explicitly use a psychological space defined using cosine distance. This concern is partly addressed by using alternative variational methods for estimating lossiness. Nevertheless, the question remains: what does a compression framework add beyond the simpler cosine dissimilarity metric of the separability of representations? Compression provides a process model where the lossiness of compression is intrinsically linked to the orthogonalization (cosine dissimilarity) of representations, while also yielding novel testable predictions about compression that extend beyond simpler dissimilarity metrics of overlap. In future work, a diffusion process that stochastically samples from latent manifolds could ground this process in retrieval and memory reconstruction processes~\cite{giguere2013limits, wolfe2021guided, banavar2024response}. Moreover, approaches adapted from compression models of visual working memory could be explored~\cite{brady2009compression, sims2015cost}. Second, it is unclear how general this framework is because our experiment focused on single-shot encoding of concrete, everyday objects. With a basis in efficient coding, we expect similar results with naturalistic images as they contain statistics that the brain is attuned to efficiently process via redundancy reduction~\cite{simoncelli2001natural, karklin2011efficient}. It would also be interesting to test pattern separation performance on sketches, drawings that focus on the gists of objects retrieved from memory. Noted as early as the origins of efficient coding theory~\cite{attneave1954some}, sketches drawn with a limited number of strokes have little information to compress, yet contain strokes that are essential for object recognition and should be retained while non-essential strokes should be discarded~\cite{mukherjee2023evaluating, long2024parallel}. Third, our analysis is limited by the temporal and spatial scale of fMRI data. For example, dentate gyrus and CA3 are not separable at this resolution. Furthermore, while we took statistical measures to limit the confounding effect of differing sizes of brain regions of interest on our metrics, measures of dimensionality remain affected by spatiotemporal scale, motivating the usage of scale-dependent measures~\cite{recanatesi2022scale}. Fourth and finally, all of our models encoded feature representations in fewer dimensions than the raw number of dimensions of the inputs (number and color of pixels) but the intrinsic dimensionality of the dataset is far lower due to redundancies among images and the predefined multi-dimensional variation between targets and lures. A stronger test of the compression and expansion hypotheses could involve determining the intrinsic dimensionality of our task's image dataset using similarity judgment tasks and computational approaches~\cite{cohen2020separability, hebart2020revealing, farrell2022gradient, fu2023dreamsim}. The usage of information bottleneck methods can elucidate how learning and forgetting different parts of inputs can support pattern separation by compressing key factors of variation across a greater number of image repetitions or experimental sessions~\cite{kim2014pruning, wanjia2021abrupt, allen2022massive, murphy2024information}.

\section{Method}

\subsection*{Experimental task}

\subsubsection*{Mnemonic similarity task} Participants were instructed to judge object images as indoor or outdoor during 128 study trials  (\textbf{Figure \ref{fig1}A}). This study phase is intended to have participants incidentally encode the stimuli in memory while performing the indoor and outdoor cover task. Next, during a test phase, participants judge stimuli as ``new,'' ``similar,'' or ``old,'' when the presented stimulus was already seen, never seen, or slightly altered (repeat, foil, and lure trials, respectively) over 192 test trials. Several performance metrics can be calculated that index different memory processes. Mnemonic discrimination performance is measured as the lure discrimination index, or the proportion of correct rejections adjusted by a response bias: $p(\textrm{``similar''}|\textrm{lure})-p(\textrm{``similar''}|\textrm{foil})$. The lure discrimination index in an independent sample was used to discretize the ``mnemonic dissimilarity'' of original and lure images into 5 bins of ease, where 5 is the easiest and most dissimilar and 1 is the hardest and most similar (\textbf{Figure \ref{fig1}B}). A recognition score is defined as correct detections corrected by the false alarm rate: $p(\textrm{``old''}|\textrm{repeat})-p(\textrm{``old''}|\textrm{foil})$. We primarily use the recognition score for quality control purposes, as we focus here on pattern separation whereas this score centers pattern completion processes~\cite{kim2013assessing}. Finally, lure false alarms are defined as as the proportion of trials where lures were mistaken as a repeat corrected for a response bias: $p(\textrm{``old''}|\textrm{lure})-p(\textrm{``old''}|\textrm{foil})$. This task allows us to probe pattern separation ease, performance, and errors.

\subsubsection*{Data and pre-processing}
We analyzed four behavioral datasets. Two datasets covered a wide range of ages: a sample of participants across the lifespan ($n=$366; $46 \pm 19$ years old; 218 women, 144 men, 4 other)~\cite{noh2021multi} and a childhood and adolescent sample ($n=$92; $16 \pm 5$ years old; 54 women, 37 men). The lifespan datasest was recruited from Amazon Mechanical Turk. Child data was recruited from Hartley Lab Participant database. Participants were recruited from the Hartley lab database which includes individuals recruited through ads on social media (e.g., Facebook and Instagram), word of mouth, local science fairs, and flyers on New York University's campus. Participants who had not previously completed an in-person study with the lab completed a brief Zoom call with a researcher. During this call, participants (and their parent or guardian, if the potential participant is under 18 years) were required to be on camera and confirm their full name and date of birth. Adult participants and parents of child and adolescent participants were additionally required to show photo identification. We also assessed two datasets of undergraduates recruited in a university setting: a sample of undergraduate students who participated for course credit ($n=$208)~\cite{stark2023optimizing} and a longitudinal sample ($n=$78) collected immediately and after 1 week. Participants in the longitudinal sample completed two study and two test phases, counterbalanced across two sessions. Half of the participants completed the test immediately after study, followed by the second study phase in the same session. They then took their second test one week later. The other half of the participants completed only the study in the first session and then completed the test one week after. They then completed the second study and second test during the second session. 

We performed several quality control steps to arrive at our final sample. First, we performed trial-level quality control. If a response was faster than 300 milliseconds or slower than 3 seconds, then the trial was excluded. Next, we performed participant-level quality control. If the participant missed more than 20\% of trials, the participant was excluded. Participants also were required to meet minimum scores on the lure discrimination index and recognition metrics indicating minimal engagement with the task. A minimum lure discrimination index of 0 was required; this minimum occurs when response bias was equal to correct rejections indicating chance-level performance. A minimum recognition score of 0.5 was required; this calculated as the correct detection minus a response bias, where chance performance is again 0. For the longitudinal dataset, we only applied the performance thresholds to exclude participants based on their immediate test.

\subsubsection*{fMRI mnemonic similarity task}
Participants (n=48; $22.9 \pm 3.6$ years old, 27 female, 21 male) completed a continuous recognition version of the mnemonic discrimination task~\cite{kirwan2007overcoming, stark2019mnemonic}; see previous publications using this dataset for more details~\cite{nash2021pattern}. A series of images of everyday objects were shown in sequence. The image set differs and predates those from those published in the prior online implementations of the task, which were used in our behavioral data above~\cite{stark2021playing}. Participants judged whether each image was new (foil trials), similar (lure trials), or old (repeat trials). The delay between the first and repeated presentation of lure or repeated objects varied with a mean lag of 19 trials. Six blocks of 107 stimuli (total of 642 images), were presented. Each block contained 32 first presentations, 16 targets, 16 similar lures, and 43 unrelated foils. Participants were required to respond within 3.0 s, after which time the stimulus was replaced by a blank screen with a fixation cross for .5 s followed by the next stimulus.

\subsection*{Neural network feature representations} 

All models were implemented using PyTorch. Autoencoder models were trained on 2,252 images across 6 image sets used in the mnemonic discrimination task. Images were originally 400x400 (x3 RGB color channels) and were downsampled to rescale according to smaller input dimensions required by the first layer of each neural network and to reduce computational costs of training. Images were normalized by the mean RGB value across all images to stabilize optimization and improves generalization. Image classification and image to text models were pre-trained on large datasets, described below. Given a stimulus from the mnemonic similarity task, the activations of the hidden layer in the autoencoders were used as the learned feature representation per image, while the activation of the convolutional or fully connected layers were used as feature representations or low-level to high-level perceptual and semantic features in the pre-trained models. The dimensionality of the layers are described in more detail below.

\subsubsection*{Image reconstruction} We use a convolutional autoencoder to learn compressed memory representations of color images in an undercomplete latent space. Stimuli were rescaled to 32x32 (x3 RGB color channels). The encoder progressively reduces the input image to a compact latent representation using three convolutional layers with stride-based downsampling and GELU activations. The tensor produced by the last convolutional layer is inputted to a fully connected bottleneck layer flattening the tensor to a 256-dimensional latent space. The decoder uses transposed convolutions to upsample the latent representation back to the original spatial dimensions, with ReLU activations and a final Tanh activation to normalize the output. We used a reconstruction loss defined as the mean squared error and trained for 250 epochs with a batch size of 32 using the Adam optimizer with a learning rate of 0.001. We refer to the 256-dimensional latent representation of each image as perceptual features.

We also use a $\beta$-VAE to learn a compressed latent representation $z$ of the input images $x$ ~\cite{kingma2013auto, higgins2017betavae}. These neural networks learn a probabilistic generative model showing how the input data $x$ depends on unobserved latent variables $z$ and approximate the optimal posterior distribution $q_\phi(z|x)$ over the latent factors given the observations. The objective of a VAE is to minimize the distance (KL-divergence) between the approximate $q_\phi(z|x)$  and true posterior distribution $p(z|x)$ by maximizing the evidence lower bound, described in the next section. The latent variables $z$ can also be used to reconstruct inputs $x$ with a measurable error. The KL-divergence and reconstruction error term in  combination forms the training objective of a variant of the VAE called the $\beta$-VAE: minimizing reconstruction error (binary cross-entropy) and minimizing the KL-divergence term to varying degrees according to the scalar parameter $\beta$. The objective function is:

\begin{equation}
L_{\beta\text{-VAE}} = \text{reconstruction error} + \beta \, \text{KL}(q_\phi(z|x) \parallel p(z))
\label{eq:L_VAE}
\end{equation}
Increasing $\beta$ enforces stronger constraints on the KL-divergence; in practice, this results in greater regularization of the latent space. In sparse $\beta$-VAEs, we further constrain the loss function to penalize a L1 regularization term on the magnitude of latent representations $z$.

The encoder consisted of four convolutional layers with ReLU activations, progressively reducing spatial dimensions while increasing feature depth. A fully connected bottleneck layer $z$ with 512 dimensions parameterized the mean and covariance of the latent distribution, and a latent vector was sampled using the reparameterization trick using a Gaussian distribution parameterized by $\phi$. The bottleneck layer factorizes the distribution over inputs $x$, forming a probabilistic representation that can disentangle the primary factors of variation across observations. The decoder, composed of transposed convolutional layers with ReLU activations, reconstructed images from the latent space. The final layer of the decoder applied a sigmoid activation to ensure output pixel values remained between 0 and 1. We trained the models over 5 random seeds using a batch size of 128 over 200 epochs with the Adam optimizer with a learning rate of 0.001. Each of the 5 seeds performed a sweep across 22 $\beta$ values ranging from $10^{-6}$ to 10 to estimate the rate-distortion function of each image.

\subsubsection*{Image classification}

We used two pre-trained convolutional neural networks called AlexNet and VGG-16 to encode different feature representations of the task stimuli. We refer to these representations as perceptual and semantic features. Stimuli were rescaled to 214x214 (x3 RGB color channels). Both networks were trained on a large number of images across a diverse set of categories in the ImageNet dataset, consisting of 1,281,167 training, 50,000 validation, and 100,000 test images across 1,000 semantic categories. The networks were trained to categorize images into the 1000 semantic categories.

AlexNet consists of nine layers. Briefly, the input layer maps unit activation to colors. Next, the output is processed sequentially by five convolutional layers, containing spatial filters that convolve with the input from the previous layer to produce the 3D tensor representing spatial activations of a particular visual pattern (see \textbf{Figure ~\ref{fig4}A}). Similar to the brain's ventral visual pathway, the hierarchical structure of convolutional layers are thought to extract low-level spatial details first then more and more abstract higher-level semantic representations in each subsequent layer to support image categorization. Next, there are three fully connected layers which flatten the 3D tensor into a vector.

For each image in the mnemonic similarity task, we obtained six vectors which we refer to as perceptual and semantic feature representations. We obtained a 4096-dimensional vector from the penultimate fully connected layer that the following layer uses for semantic categorization. The penultimate layer is used instead of the last fully connected layer in order to obtain more generalizable, less task-specific, and richer features that are useful for transfer learning, clustering, and similarity tasks. In contrast, the last layer's features are tailored for categorizing images into 1,000 categories. We also obtained five feature representations with differing dimensionality from the activation of each convolutional layer. The feature representations were 96-dimensional in the first layer, 256-dimensional in the second layer, 384-dimensional in the third layer, 384-dimensional in the fourth layer, and 256-dimensional in the fifth layer.

VGG-16 is another convolutional neural network trained to categorize images on the ImageNet dataset but has a different architecture. The input dimensionality for images is again 224x224x3 and the network is again trained to classify images into 1000 categories. The input image is processed by 13 convolutional layers and 3 fully connected layers. For each image in the mnemonic similarity task, we obtained a 4096-dimensional feature representation from the penultimate fully connected layer.

\subsubsection*{Image to text} 

To obtain semantic feature representations, we used the contrastive language-image pre-training (CLIP) neural network framework with the ViT-H/14 image embedding variant, a hierarchical vision transformer. This model was pre-trained with the LAION-2B English dataset, a dataset of 2 billion English image-text pairs. Briefly, a vision encoder transforms images into vector representations, using pixel patches. The text encoder processes text descriptions through a transformer architecture, creating word-vector embeddings in the same semantic space as the visual features. The model is trained to optimize a contrastive objective, maximizing similarity between paired image-text embeddings while minimizing similarity with unpaired samples. This objective constrains the vision and text encoders to learn representations in a joint embedding space where semantically similar concepts are more similarly represented regardless of whether their source was an image or text. The resulting image embeddings from this model can be used for zero-shot image classification and other tasks without requiring task-specific fine-tuning. For example, the model performs zero-shot categorization of the Imagenet dataset with a 78.0\% accuracy despite not being explicitly trained to do so. For each image stimulus in the mnemonic similarity task, we use this pre-trained model to obtain a text description. This description is tokenized and inputted to the pre-trained text encoder to extract a 768-dimensional word-vector embedding. We refer to the 768-dimensional vectors as semantic features.

\subsection*{Calculating lossiness}

\subsubsection*{Lossiness from a simulation of memory retrieval and identification}
To compute lossiness for all models, except for the $\beta$-VAE, we adapted a previously published information theoretic algorithm~\cite{sims2018efficient}. The algorithm infers a loss function based on efficiently coding (compressing) an input $X_1$ into an output $X_2$. In that work, the lossiness of compression explained when two perceptually similar items are confused (generalized) or discriminated across many sensory modalities, such as a range of tones, colors, line lengths, and shapes.

In the context of the mnemonic similarity task, we determine the perceptual loss when compressing $X_{targets}$ and $X_{lures}$ in latent space. To simulate the cost of remembering $X_{targets}$ given $X_{lures}$, we linearly interpolate between $x_{target}$ and $x_{lure}$ in latent space. Similar images are thought to exist on a low-dimensional manifold in this latent space \cite{dicarlo_how_2012, CHUNG2021137}, and linearly interpolating between points on this manifold can be thought of as a trajectory along the manifold's surface from one image to another. This trajectory represents samples that transform according to a primary factor of variation, such as the rotation leftward versus rightward a face, the shape of a square to circle, and the color of a blue to orange object. After generating intermediate representations along this manifold, we calculate the cosine distance between all pairs of latent representations. Then, we convert this distance matrix to a similarity matrix and normalize values by the the maximum similarity across latent representations. Finally we simulate a confusion matrix that represents the perceptual information channel, where rows are input representations and columns are output representations, based by converting all values of the similarity matrix to probablities marginalizing over the output columns. Drawing 250 samples from this probability matrix simulates how often an input representation along the target-lure image manifold is mistaken for each output representation. Previously published R code was used to fit a model based on rate-distortion theory to the confusion matrix. Specifically, the lossiness was inferred using the Blahut algorithm for computing an optimal, capacity-limited information channel. This optimization procedure is grounded in rate-distortion theory and iteratively adjusts conditional probabilities to minimize the lossiness at a given information rate. Model fitting was performed over 100 iterations, ensuring convergence to a stable solution, by maximizing the log-likelihood of the observed confusion matrix under the inferred probability distributions, incorporating a prior over the lossiness that penalized deviations from symmetry and constrained diagonal elements. To improve robustness, multiple fitting attempts were conducted, selecting the best fit based on the highest log-likelihood value. This procedure returns a cost matrix where each element is the lossiness or cost of error for confusing an input (row) with each output (column element). The lossiness used in the current paper is the average lossiness across the cost matrix per pair of target and lure images.

\subsubsection*{Optimal lossiness from variational methods}

In a $\beta$-VAE, the loss function is defined with a trade-off that parallels rate-distortion trade-offs. Recall that the objective function in Equation \ref{eq:L_VAE} is $L_{\beta\text{-VAE}} = \text{reconstruction error} + \beta \, \text{KL}(q_\phi(z|x) \parallel p(z))$. Rate-distortion functions can be approximated by interpreting the reconstruction error (here defined as the binary cross entropy loss) as the distortion $D$, the information-limiting KL-divergence term as the rate $R$, and $\beta$ as a Lagrange multiplier that scales the regularization. Larger $\beta$ enforces more constraints on the information channel.

This objective is a special unsupervised case of the supervised information bottleneck method which separates task-relevant and irrelevant information~\cite{alemi2017deep}. Formulating the objective function in this way enables a controllable rate-distortion trade-off:

\begin{equation}
L_{\text{rate-distortion}} = D + \beta \, R.
\end{equation}

This rate-distortion loss function characterizes how a better lossy compression minimizes both the distortion and the information rate needed to transmit the source. Here, increasing $\beta$ enforces more conservative information rates, or more aggressive compression characterizing a more capacity-limited information channel. Using this method, the lossiness of each image was measured as the average reconstruction error of the target and lure images across $\beta$ for the seed with the lowest train/test loss. The normalized rate was measured by calculating the slope from a linear function that best fit the rate-distortion function plotted in a semi-log plot: $\frac{log_{10}(R)}{D}$.

\subsection*{MRI data acquisition}
Neuroimaging was performed on a Siemens 3 Tesla TIM Trio scanner using a 32-channel receive-only head coil. Structural images were acquired using a T1-weighted MP-RAGE sequence with the following parameters: 192 interleaved slices; total acquisition = 8:55 min; TR = 20 ms; TE = 4.92 ms; flip angle = 25°; field of view = 256 mm; slice thickness = 1 mm; voxel resolution = 1.0 × 1.0 × 1.0 mm; 1 average. A T2-weighted pulse sequence was acquired with the following parameters: 35 interleaved slices; total acquisition = 9:44 min; TR = 6,000 ms; TE = 64 ms; flip angle = 129°; field of view = 200 mm; slice thickness = 2 mm; voxel resolution = 0.4 × 0.4 × 2.0 mm; averages = 2. This scan was positioned perpendicular to the longitudinal axis of the hippocampus for each participant prior to acquisition.

High-resolution functional images were acquired using gradient-echo echoplanar, T2*-weighted pulse sequence utilizing a multiband (MB) technique~\cite{xu2013evaluation}; six functional scans were conducted that coincided with six blocks of the task. These scans had the following parameters: 72 interleaved slices; TA = 6:25 min; TR = 875 ms; TE = 43.6 ms; flip angle 55°; field of view = 180 mm; slice thickness = 1.8 mm; voxel resolution = 1.8 × 1.8 × 1.8 mm; MB factor = 8; measurements = 428. These scans were aligned with the longitudinal axis of the hippocampus for each participant. The first four TRs of each run were discarded to allow for T1 equilibration.

\subsection*{fMRI preprocessing}

MRI data were analyzed using the Analysis of Functional NeuroImages (AFNI) software (version 22.1.09;~\cite{cox1996afni}) following the first steps of a prior publication using the same data~\cite{nash2021pattern}. Given the short acquisition time, functional scans were not slice-time corrected. Motion correction for functional scans was calculated to align each volume to the single volume of the experiment with the smallest number of outlier voxel values. Structural scans were also aligned with the minimum-outlier functional volume. Rotated structural scans were then skull-stripped and warped into MNI space using a nonlinear diffeomorphic transformation. The motion correction and MNI normalization spatial transformations were concatenated and applied to the functional scans in a single step, thus resulting in a single spatial transformation for functional data. Functional data were scaled by the mean of the overall signal for each run. No blurring was done to functional data as part of preprocessing and spatial resolution of functional scans was maintained at 1.8 $\text{mm}^3$. 

Using this minimally pre-processed data, we performed trial-wise GLM estimation using the GLMsingle toolbox implemented in Python~\cite{prince2022improving}. This toolbox identifies an optimal HRF from a library of 20 canonical functional forms, performs automated denoising using data-driven nuisance regressors from selected repeat trials, and implements fractional-ridge regularization to regularize estimates based on voxel-wise reliabilities~\cite{rokem2020fractional}. This pipeline does not benefit from regressing out putative nuisance components of fMRI data, such as motion, white matter, or cerebrospinal fluid, prior to running GLMsingle because those steps can bias the data-driven learning of nuisance components which potentially overlap with signal components of interest. For these reasons, our pre-processing departed from the steps in the prior work using this dataset~\cite{nash2021pattern}, i.e., we also did not use the motion scrubbing (framewise censoring of TRs with greater than 0.3° of rotation or 0.6 mm of translation in any direction, as well as the immediately preceding TR) or coverage masks removing very low EPI signal.  

Region of interest maps were created for the ventral visual stream (V1, V2, V3, V4, and IT) that progressively extract low-level spatial details and high-level semantic abstractions, as well as the hippocampal subregions (anterolateral entorhinal cortex, DG/CA3, CA1, and subiculum) that classically support pattern separation for inputs from the object processing stream. 

For each subject, we masked all ROIs. Then, for 1 of the 6 runs, we split the 2nd run for each participant into two half runs because GLMsingle requires repeat trials across runs for cross-validation and the original task only includes repeat trials within a run. With stimulus duration = 2.5 s and repetition time = 0.875 s, we ran GLMsingle with default options. Outputs used for further analysis included the per-trial evoked activity $\beta$ maps and map of HRF indices. Split-half reliability on $\beta$s were computed voxelwise, though we did not exclude any low-reliability voxels to avoid potential double dipping, biasing selection for voxels showing stronger memory reactivation for repeats.

\subsection*{fMRI analysis}

To analyze representations of distinct stimuli, we used the trial-wise evoked activity magnitudes $\beta$ as well as a data-driven selection of a hemodynamic response function (HRF) from a library of canonical functions. We use the magnitude and HRF to construct the activity time series per voxel evoked by each stimulus. The metrics calculated using the neural data scale strongly with the size of the input. Therefore, to obtain regional metrics which are more fairly comparable across ROIs of substantially different size (e.g. larger primary visual cortex versus smaller dentate gyrus/CA3), we sampled 100 voxel-wise time series over 100 iterations to compute averaged regional signatures of lossy compression. 

\subsubsection*{Neural dimensionality of target and lure representations} The first signature of lossy compression we investigated treats dimensionality reduction as a form of lossy compression, because reconstructing an input using a reduced number of dimensions introduces distortions. The key quantity is the representational dimensionality of target-lure stimuli, calculated as the participation ratio of the covariance matrix of the target and lure time series. 

\begin{equation}
\text{Dimensionality} = \frac{\left( \sum_{i=1}^{N} \lambda_i \right)^2}{\sum_{i=1}^{N} \lambda_i^2}
\end{equation}

\noindent where $\lambda_i$ are the eigenvalues of the covariance matrix of the target and lure time series, and $N$ is the number of dimensions. Participation ratios are like a continuous and normalized version of the discrete principal components that explain all variance in the data. Higher participation ratios indicate neural representations with greater dimensionality and a larger capacity to represent more features or stimuli~\cite{rigotti2013importance, gao2017theory, zhou2022compression} which may support the separability of distinct inputs or input categories~\cite{cayco-gajic_re-evaluating_2019, CHUNG2021137}. In contrast, a lossy compression account of pattern separation predicts that separability is supposed by lower dimensionality, selectively discarding less relevant features to only retain essential information. 

A signature of lossy compression is therefore dimensionality reduction for correct versus incorrect trials: the difference between the dimensionality of trials with correct lure responses ('similar') minus the dimensionality of trials with incorrect lure false alarm responses ('old'). 

\begin{equation}
\Delta \text{Dimensionality} = \text{Dimensionality}_{\text{correct}} - \text{Dimensionality}_{\text{incorrect}}
\end{equation}

\noindent where $\text{Dimensionality}_{\text{correct}}$ is the participation ratio of lure trials with correct “similar” responses, and $\text{Dimensionality}_{\text{incorrect}}$ is the participation ratio of lure trials with incorrect “old” responses.

\subsubsection*{Information rate between target and lure representations}
The second signature of lossy compression we investigated is the mutual information between the neural representations of targets and lures. The mutual information is the amount of information retained between representations and should be lower for correct trials versus incorrect trials according to a lossy compression account. Mutual information was calculated by discretizing values into fixed bins and using the mutual information score between the resulting discretized distributions~\cite{scikitlearn}:

\begin{equation}
I(X;Y) = \sum_{i=1}^{|X|} \sum_{j=1}^{|Y|} 
\frac{n_{ij}}{n} \, \log \left( \frac{n_{ij} \cdot n}{n_{i\cdot} \, n_{\cdot j}} \right)
\end{equation}

\noindent where $n_{ij}$ is the number of co-occurrences of bin $i$ in $X$ and bin $j$ in $Y$, 
$n_{i\cdot}$ and $n_{\cdot j}$ are the corresponding marginal counts, 
and $n$ is the total number of samples. To calculate within-region information rates for incorrect trials and correct trials, we concatenated and flattened all voxel-wise time courses for targets and separately for lures. Then, we separated these lure and target time courses by whether the trial was evaluated as a correct lure response ('similar') and incorrect lure false alarm response ('old'). Finally, for correct trial mutual information, we defined X as the target time courses for the correct trials and Y as the lure time courses for the correct trials. For incorrect trial mutual information, we did the same but with only the incorrect trials. The direction of results did not differ when using more advanced adaptive discretization algorithms which are better able to capture non-linear relationships~\cite{kraskov2004estimating, lizier2014jidt}.

\subsection*{Statistical analysis}

To test the effect of lossiness on the ease of pattern separation performance, we used a Spearman correlation between lure bin and lossiness. To test the effect of lossiness on pattern separation performance, we tested a linear model with mixed effects at the trial-level nested by participant:

\begin{equation}
    \text{Lure discrimination index} = \beta_0 + \beta_1 \cdot \text{lossiness} + \beta_2 \cdot \text{Age} + (1|\text{participant}) + \epsilon
\end{equation}

The lure discrimination index was calculated across trials within each lure bin. The lossiness was the mean across the images presented in each lure bin. We only included age as a covariate for the dataset that spanned the lifespan and the dataset containing children and adolescents. Finally, the effects of age are often non-linear. To test both linear and non-linear effects of age on the relationship between lossiness and pattern separation performance, we used generalized additive models with penalized splines, a method which allows for statistically rigorous modeling of linear and nonlinear effects while minimizing over-fitting~\cite{wood2004stable}. We tested the model:

\begin{equation}
    \text{Lure discrimination index} = \beta_0 + s(\text{Lossiness, by age}, k=4) + \epsilon
\end{equation}

The model included $k$ as a smooth term for lossiness, capturing non-linear effects of lossiness across different age groups, using 4 basis functions and fit using restricted maximum likelihood estimation and fixed effects.

\subsection*{Citation diversity statement}

Recent work in several fields of science has identified a bias in citation practices such that papers from women and other minority scholars are under-cited relative to the number of such papers in the field \cite{mitchell2013gendered,dion2018gendered,caplar2017quantitative, maliniak2013gender, Dworkin2020.01.03.894378, bertolero2021racial, wang2021gendered, chatterjee2021gender, fulvio2021imbalance}. Here we sought to proactively consider choosing references that reflect the diversity of the field in thought, form of contribution, gender, race, ethnicity, and other factors. First, we obtained the predicted gender of the first and last author of each reference by using databases that store the probability of a first name being carried by a woman \cite{Dworkin2020.01.03.894378,zhou_dale_2020_3672110}. By this measure (and excluding self-citations to the first and last authors of our current paper), our references contain 12.27\% woman(first)/woman(last), 12.63\% man/woman, 21.27\% woman/man, and 53.84\% man/man. This method is limited in that a) names, pronouns, and social media profiles used to construct the databases may not, in every case, be indicative of gender identity and b) it cannot account for intersex, non-binary, or transgender people. Second, we obtained predicted racial/ethnic category of the first and last author of each reference by databases that store the probability of a first and last name being carried by an author of color \cite{ambekar2009name, sood2018predicting}. By this measure (and excluding self-citations), our references contain 6.18\% author of color (first)/author of color(last), 12.90\% white author/author of color, 22.74\% author of color/white author, and 58.18\% white author/white author. This method is limited in that a) names and Florida Voter Data to make the predictions may not be indicative of racial/ethnic identity, and b) it cannot account for Indigenous and mixed-race authors, or those who may face differential biases due to the ambiguous racialization or ethnicization of their names. We look forward to future work that could help us to better understand how to support equitable practices in science.

\section{Acknowledgments}

D.Z. acknowledges funding from the George E. Hewitt Foundation for Medical Research. A.M.B. acknowledges funding from the National Institute on Aging (R21AG072673 and R01AG088306). M.Y. and A.M.B. acknowledge funding from the National Institute for Mental Health (R01MH128306).

\section{Data availability}
All fMRI data are available at \href{https://openneuro.org/datasets/ds002168}{OpenNeuro} and pre-processing scripts used in data analyses are available at \href{https://github.com/Kirwanlab/MST_WholeBrain}{GitHub}. 

\bibliographystyle{ieeetr}

\bibliography{ccn_style}

\begin{thebibliography}{100}

\bibitem{noh2014multilevel}
S.~M. Noh, V.~X. Yan, M.~S. Vendetti, A.~D. Castel, and R.~A. Bjork, ``Multilevel induction of categories: Venomous snakes hijack the learning of lower category levels,'' {\em Psychological science}, vol.~25, no.~8, pp.~1592--1599, 2014.

\bibitem{botvinick2015reinforcement}
M.~Botvinick, A.~Weinstein, A.~Solway, and A.~Barto, ``Reinforcement learning, efficient coding, and the statistics of natural tasks,'' {\em Current opinion in behavioral sciences}, vol.~5, pp.~71--77, 2015.

\bibitem{bornstein2023associative}
A.~M. Bornstein, M.~Aly, S.~F. Feng, N.~B. Turk-Browne, K.~A. Norman, and J.~D. Cohen, ``Associative memory retrieval modulates upcoming perceptual decisions,'' {\em Cognitive, Affective, \& Behavioral Neuroscience}, vol.~23, no.~3, pp.~645--665, 2023.

\bibitem{pettine2023human}
W.~W. Pettine, D.~V. Raman, A.~D. Redish, and J.~D. Murray, ``Human generalization of internal representations through prototype learning with goal-directed attention,'' {\em Nature human behaviour}, vol.~7, no.~3, pp.~442--463, 2023.

\bibitem{cayco-gajic_re-evaluating_2019}
N.~A. Cayco-Gajic and R.~A. Silver, ``Re-evaluating {Circuit} {Mechanisms} {Underlying} {Pattern} {Separation},'' {\em Neuron}, vol.~101, pp.~584--602, Feb. 2019.
\newblock Publisher: Elsevier.

\bibitem{hulbert2015neural}
J.~C. Hulbert and K.~Norman, ``Neural differentiation tracks improved recall of competing memories following interleaved study and retrieval practice,'' {\em Cerebral Cortex}, vol.~25, no.~10, pp.~3994--4008, 2015.

\bibitem{kim2017neural}
G.~Kim, K.~A. Norman, and N.~B. Turk-Browne, ``Neural differentiation of incorrectly predicted memories,'' {\em Journal of Neuroscience}, vol.~37, no.~8, pp.~2022--2031, 2017.

\bibitem{khoudary2022precision}
A.~Khoudary, M.~A. Peters, and A.~M. Bornstein, ``Precision-weighted evidence integration predicts time-varying influence of memory on perceptual decisions,'' {\em Cognitive Computational Neuroscience}, 2022.

\bibitem{noh2023memory}
S.~M. Noh, U.~K. Singla, I.~J. Bennett, and A.~M. Bornstein, ``Memory precision and age differentially predict the use of decision-making strategies across the lifespan,'' {\em Scientific Reports}, vol.~13, no.~1, p.~17014, 2023.

\bibitem{bein2023predictions}
O.~Bein, C.~Gasser, T.~Amer, A.~Maril, and L.~Davachi, ``Predictions transform memories: How expected versus unexpected events are integrated or separated in memory,'' {\em Neuroscience \& Biobehavioral Reviews}, vol.~153, p.~105368, 2023.

\bibitem{bartlett1995remembering}
F.~C. Bartlett, {\em Remembering: A study in experimental and social psychology}.
\newblock Cambridge university press, 1995.

\bibitem{loftus1996memory}
E.~F. Loftus, ``Memory distortion and false memory creation,'' {\em Journal of the American Academy of Psychiatry and the Law Online}, vol.~24, no.~3, pp.~281--295, 1996.

\bibitem{yassa2011pattern}
M.~A. Yassa, J.~W. Lacy, S.~M. Stark, M.~S. Albert, M.~Gallagher, and C.~E. Stark, ``Pattern separation deficits associated with increased hippocampal ca3 and dentate gyrus activity in nondemented older adults,'' {\em Hippocampus}, vol.~21, no.~9, pp.~968--979, 2011.

\bibitem{motley2012parametric}
S.~E. Motley and C.~B. Kirwan, ``A parametric investigation of pattern separation processes in the medial temporal lobe,'' {\em Journal of Neuroscience}, vol.~32, no.~38, pp.~13076--13084, 2012.

\bibitem{bakker2008pattern}
A.~Bakker, C.~B. Kirwan, M.~Miller, and C.~E. Stark, ``Pattern separation in the human hippocampal ca3 and dentate gyrus,'' {\em science}, vol.~319, no.~5870, pp.~1640--1642, 2008.

\bibitem{leutgeb2007pattern}
J.~K. Leutgeb, S.~Leutgeb, M.-B. Moser, and E.~I. Moser, ``Pattern separation in the dentate gyrus and ca3 of the hippocampus,'' {\em science}, vol.~315, no.~5814, pp.~961--966, 2007.

\bibitem{marr1991simple}
D.~Marr, D.~Willshaw, and B.~McNaughton, {\em Simple memory: a theory for archicortex}.
\newblock Springer, 1991.

\bibitem{o1994hippocampal}
R.~C. O'Reilly and J.~L. McClelland, ``Hippocampal conjunctive encoding, storage, and recall: Avoiding a trade-off,'' {\em Hippocampus}, vol.~4, no.~6, pp.~661--682, 1994.

\bibitem{mcclelland1996considerations}
J.~L. McClelland and N.~H. Goddard, ``Considerations arising from a complementary learning systems perspective on hippocampus and neocortex,'' {\em Hippocampus}, vol.~6, no.~6, pp.~654--665, 1996.

\bibitem{o2001conjunctive}
R.~C. O'Reilly and J.~W. Rudy, ``Conjunctive representations in learning and memory: principles of cortical and hippocampal function.,'' {\em Psychological review}, vol.~108, no.~2, p.~311, 2001.

\bibitem{o2002hippocampal}
R.~C. O'Reilly and K.~A. Norman, ``Hippocampal and neocortical contributions to memory: Advances in the complementary learning systems framework,'' {\em Trends in cognitive sciences}, vol.~6, no.~12, pp.~505--510, 2002.

\bibitem{suthana2015specific}
N.~A. Suthana, N.~N. Parikshak, A.~D. Ekstrom, M.~J. Ison, B.~J. Knowlton, S.~Y. Bookheimer, and I.~Fried, ``Specific responses of human hippocampal neurons are associated with better memory,'' {\em Proceedings of the National Academy of Sciences}, vol.~112, no.~33, pp.~10503--10508, 2015.

\bibitem{dimsdale2018ca1}
H.~R. Dimsdale-Zucker, M.~Ritchey, A.~D. Ekstrom, A.~P. Yonelinas, and C.~Ranganath, ``Ca1 and ca3 differentially support spontaneous retrieval of episodic contexts within human hippocampal subfields,'' {\em Nature communications}, vol.~9, no.~1, p.~294, 2018.

\bibitem{sakon2019neural}
J.~J. Sakon and W.~A. Suzuki, ``A neural signature of pattern separation in the monkey hippocampus,'' {\em Proceedings of the national academy of sciences}, vol.~116, no.~19, pp.~9634--9643, 2019.

\bibitem{stark2019mnemonic}
S.~M. Stark, C.~B. Kirwan, and C.~E. Stark, ``Mnemonic similarity task: A tool for assessing hippocampal integrity,'' {\em Trends in cognitive sciences}, vol.~23, no.~11, pp.~938--951, 2019.

\bibitem{treves1994computational}
A.~Treves and E.~T. Rolls, ``Computational analysis of the role of the hippocampus in memory,'' {\em Hippocampus}, vol.~4, no.~3, pp.~374--391, 1994.

\bibitem{norman2010hippocampus}
K.~A. Norman, ``How hippocampus and cortex contribute to recognition memory: revisiting the complementary learning systems model,'' {\em Hippocampus}, vol.~20, no.~11, pp.~1217--1227, 2010.

\bibitem{leal2018integrating}
S.~L. Leal and M.~A. Yassa, ``Integrating new findings and examining clinical applications of pattern separation,'' {\em Nature neuroscience}, vol.~21, no.~2, pp.~163--173, 2018.

\bibitem{10.3389/fnbeh.2013.00096}
A.~Santoro, ``Reassessing pattern separation in the dentate gyrus,'' {\em Frontiers in Behavioral Neuroscience}, vol.~7, 2013.

\bibitem{marr1969theory}
D.~Marr, ``A theory of cerebellar cortex. 202: 437--470,'' 1969.

\bibitem{albus1971theory}
J.~S. Albus, ``A theory of cerebellar function,'' {\em Mathematical biosciences}, vol.~10, no.~1-2, pp.~25--61, 1971.

\bibitem{myers2009role}
C.~E. Myers and H.~E. Scharfman, ``A role for hilar cells in pattern separation in the dentate gyrus: a computational approach,'' {\em Hippocampus}, vol.~19, no.~4, pp.~321--337, 2009.

\bibitem{myers2011pattern}
C.~E. Myers and H.~E. Scharfman, ``Pattern separation in the dentate gyrus: a role for the ca3 backprojection,'' {\em Hippocampus}, vol.~21, no.~11, pp.~1190--1215, 2011.

\bibitem{billings_network_2014}
G.~Billings, E.~Piasini, A.~Lőrincz, Z.~Nusser, and R.~Silver, ``Network {Structure} within the {Cerebellar} {Input} {Layer} {Enables} {Lossless} {Sparse} {Encoding},'' {\em Neuron}, vol.~83, pp.~960--974, Aug. 2014.
\newblock Publisher: Elsevier.

\bibitem{chavlis2017dendrites}
S.~Chavlis, P.~C. Petrantonakis, and P.~Poirazi, ``Dendrites of dentate gyrus granule cells contribute to pattern separation by controlling sparsity,'' {\em Hippocampus}, vol.~27, no.~1, pp.~89--110, 2017.

\bibitem{guzman_how_2021}
S.~J. Guzman, A.~Schlögl, C.~Espinoza, X.~Zhang, B.~A. Suter, and P.~Jonas, ``How connectivity rules and synaptic properties shape the efficacy of pattern separation in the entorhinal cortex–dentate gyrus–{CA3} network,'' {\em Nature Computational Science}, vol.~1, pp.~830--842, Dec. 2021.

\bibitem{chanales2017overlap}
A.~J. Chanales, A.~Oza, S.~E. Favila, and B.~A. Kuhl, ``Overlap among spatial memories triggers repulsion of hippocampal representations,'' {\em Current Biology}, vol.~27, no.~15, pp.~2307--2317, 2017.

\bibitem{sims2018efficient}
C.~R. Sims, ``Efficient coding explains the universal law of generalization in human perception,'' {\em Science}, vol.~360, no.~6389, pp.~652--656, 2018.

\bibitem{mack2020ventromedial}
M.~L. Mack, A.~R. Preston, and B.~C. Love, ``Ventromedial prefrontal cortex compression during concept learning,'' {\em Nature communications}, vol.~11, no.~1, p.~46, 2020.

\bibitem{zhou2022efficient}
D.~Zhou, C.~W. Lynn, Z.~Cui, R.~Ciric, G.~L. Baum, T.~M. Moore, D.~R. Roalf, J.~A. Detre, R.~C. Gur, R.~E. Gur, {\em et~al.}, ``Efficient coding in the economics of human brain connectomics,'' {\em Network Neuroscience}, vol.~6, no.~1, pp.~234--274, 2022.

\bibitem{higgins2017betavae}
I.~Higgins, L.~Matthey, A.~Pal, C.~Burgess, X.~Glorot, M.~Botvinick, S.~Mohamed, and A.~Lerchner, ``beta-{VAE}: Learning basic visual concepts with a constrained variational framework,'' in {\em International Conference on Learning Representations}, 2017.

\bibitem{higgins2021unsupervised}
I.~Higgins, L.~Chang, V.~Langston, D.~Hassabis, C.~Summerfield, D.~Tsao, and M.~Botvinick, ``Unsupervised deep learning identifies semantic disentanglement in single inferotemporal face patch neurons,'' {\em Nature communications}, vol.~12, no.~1, p.~6456, 2021.

\bibitem{rigotti2013importance}
M.~Rigotti, O.~Barak, M.~R. Warden, X.-J. Wang, N.~D. Daw, E.~K. Miller, and S.~Fusi, ``The importance of mixed selectivity in complex cognitive tasks,'' {\em Nature}, vol.~497, no.~7451, pp.~585--590, 2013.

\bibitem{hasselmo2006role}
M.~E. Hasselmo, ``The role of acetylcholine in learning and memory,'' {\em Current opinion in neurobiology}, vol.~16, no.~6, pp.~710--715, 2006.

\bibitem{gao2015simplicity}
P.~Gao and S.~Ganguli, ``On simplicity and complexity in the brave new world of large-scale neuroscience,'' {\em Current opinion in neurobiology}, vol.~32, pp.~148--155, 2015.

\bibitem{gao2017theory}
P.~Gao, E.~Trautmann, B.~Yu, G.~Santhanam, S.~Ryu, K.~Shenoy, and S.~Ganguli, ``A theory of multineuronal dimensionality, dynamics and measurement,'' {\em BioRxiv}, p.~214262, 2017.

\bibitem{stringer2019high}
C.~Stringer, M.~Pachitariu, N.~Steinmetz, M.~Carandini, and K.~D. Harris, ``High-dimensional geometry of population responses in visual cortex,'' {\em Nature}, vol.~571, no.~7765, pp.~361--365, 2019.

\bibitem{cohen2020separability}
U.~Cohen, S.~Chung, D.~D. Lee, and H.~Sompolinsky, ``Separability and geometry of object manifolds in deep neural networks,'' {\em Nature communications}, vol.~11, no.~1, p.~746, 2020.

\bibitem{koolschijn2021memory}
R.~S. Koolschijn, A.~Shpektor, W.~T. Clarke, I.~B. Ip, D.~Dupret, U.~E. Emir, and H.~C. Barron, ``Memory recall involves a transient break in excitatory-inhibitory balance,'' {\em Elife}, vol.~10, p.~e70071, 2021.

\bibitem{10.5555/1146355}
T.~M. Cover and J.~A. Thomas, {\em Elements of Information Theory (Wiley Series in Telecommunications and Signal Processing)}.
\newblock USA: Wiley-Interscience, 2006.

\bibitem{nagy2020optimal}
D.~G. Nagy, B.~T{\"o}r{\"o}k, and G.~Orb{\'a}n, ``Optimal forgetting: Semantic compression of episodic memories,'' {\em PLoS Computational Biology}, vol.~16, no.~10, p.~e1008367, 2020.

\bibitem{nagy2025interplay}
D.~G. Nagy, G.~Orban, and C.~M. Wu, ``Interplay of episodic and semantic memory arises from adaptive compression,'' 2025.

\bibitem{shannon1959coding}
C.~E. Shannon {\em et~al.}, ``Coding theorems for a discrete source with a fidelity criterion,'' {\em IRE Nat. Conv. Rec}, vol.~4, no.~142-163, p.~1, 1959.

\bibitem{yonelinas2010recollection}
A.~P. Yonelinas, M.~Aly, W.-C. Wang, and J.~D. Koen, ``Recollection and familiarity: Examining controversial assumptions and new directions,'' {\em Hippocampus}, vol.~20, no.~11, pp.~1178--1194, 2010.

\bibitem{leal2014asymmetric}
S.~L. Leal, S.~K. Tighe, and M.~A. Yassa, ``Asymmetric effects of emotion on mnemonic interference,'' {\em Neurobiology of learning and memory}, vol.~111, pp.~41--48, 2014.

\bibitem{noh2021multi}
S.~M. Noh, K.~W. Cooper, C.~E. Stark, and A.~M. Bornstein, ``Multi-step inference can be improved across the lifespan with individualized memory interventions,'' {\em PsyArXiv}, 2024.

\bibitem{nash2021pattern}
M.~I. Nash, C.~B. Hodges, N.~M. Muncy, and C.~B. Kirwan, ``Pattern separation beyond the hippocampus: A high-resolution whole-brain investigation of mnemonic discrimination in healthy adults,'' {\em Hippocampus}, vol.~31, no.~4, pp.~408--421, 2021.

\bibitem{stark2023optimizing}
C.~E. Stark, J.~A. Noche, J.~R. Ebersberger, L.~Mayer, and S.~M. Stark, ``Optimizing the mnemonic similarity task for efficient, widespread use,'' {\em Frontiers in behavioral neuroscience}, vol.~17, p.~1080366, 2023.

\bibitem{banavar2024response}
N.~V. Banavar, S.~M. Noh, C.~N. Wahlheim, B.~S. Cassidy, C.~B. Kirwan, C.~E. Stark, and A.~M. Bornstein, ``A response time model of the three-choice mnemonic similarity task provides stable, mechanistically interpretable individual-difference measures,'' {\em Frontiers in Human Neuroscience}, vol.~18, p.~1379287, 2024.

\bibitem{NIPS2012_c399862d}
A.~Krizhevsky, I.~Sutskever, and G.~E. Hinton, ``Imagenet classification with deep convolutional neural networks,'' in {\em Advances in Neural Information Processing Systems} (F.~Pereira, C.~Burges, L.~Bottou, and K.~Weinberger, eds.), vol.~25, Curran Associates, Inc., 2012.

\bibitem{Simonyan2014VeryDC}
K.~Simonyan and A.~Zisserman, ``Very deep convolutional networks for large-scale image recognition,'' {\em CoRR}, vol.~abs/1409.1556, 2014.

\bibitem{lewis2015neural}
J.~A. Lewis-Peacock, A.~T. Drysdale, and B.~R. Postle, ``Neural evidence for the flexible control of mental representations,'' {\em Cerebral Cortex}, vol.~25, no.~10, pp.~3303--3313, 2015.

\bibitem{Radford2021LearningTV}
A.~Radford, J.~W. Kim, C.~Hallacy, A.~Ramesh, G.~Goh, S.~Agarwal, G.~Sastry, A.~Askell, P.~Mishkin, J.~Clark, G.~Krueger, and I.~Sutskever, ``Learning transferable visual models from natural language supervision,'' in {\em ICML}, 2021.

\bibitem{yamins}
D.~L.~K. Yamins, H.~Hong, C.~F. Cadieu, E.~A. Solomon, D.~Seibert, and J.~J. DiCarlo, ``Performance-optimized hierarchical models predict neural responses in higher visual cortex,'' {\em Proceedings of the National Academy of Sciences}, vol.~111, no.~23, pp.~8619--8624, 2014.

\bibitem{benna2021place}
M.~K. Benna and S.~Fusi, ``Place cells may simply be memory cells: Memory compression leads to spatial tuning and history dependence,'' {\em Proceedings of the National Academy of Sciences}, vol.~118, no.~51, p.~e2018422118, 2021.

\bibitem{wang_better_2023}
A.~Y. Wang, K.~Kay, T.~Naselaris, M.~J. Tarr, and L.~Wehbe, ``Better models of human high-level visual cortex emerge from natural language supervision with a large and diverse dataset,'' {\em Nature Machine Intelligence}, vol.~5, pp.~1415--1426, Dec. 2023.

\bibitem{recanatesi2019dimensionality}
S.~Recanatesi, M.~Farrell, M.~Advani, T.~Moore, G.~Lajoie, and E.~Shea-Brown, ``Dimensionality compression and expansion in deep neural networks,'' {\em arXiv preprint arXiv:1906.00443}, 2019.

\bibitem{farrell2022gradient}
M.~Farrell, S.~Recanatesi, T.~Moore, G.~Lajoie, and E.~Shea-Brown, ``Gradient-based learning drives robust representations in recurrent neural networks by balancing compression and expansion,'' {\em Nature Machine Intelligence}, vol.~4, no.~6, pp.~564--573, 2022.

\bibitem{alemi2017deep}
A.~A. Alemi, I.~Fischer, J.~V. Dillon, and K.~Murphy, ``Deep variational information bottleneck,'' in {\em International Conference on Learning Representations}, 2017.

\bibitem{bates2020efficient}
C.~J. Bates and R.~A. Jacobs, ``Efficient data compression in perception and perceptual memory.,'' {\em Psychological review}, vol.~127, no.~5, p.~891, 2020.

\bibitem{zhou2022compression}
D.~Zhou, J.~Z. Kim, A.~R. Pines, V.~J. Sydnor, D.~R. Roalf, J.~A. Detre, R.~C. Gur, R.~E. Gur, T.~D. Satterthwaite, and D.~S. Bassett, ``Compression supports low-dimensional representations of behavior across neural circuits,'' {\em arXiv preprint arXiv:2211.16599}, 2022.

\bibitem{lacy2011distinct}
J.~W. Lacy, M.~A. Yassa, S.~M. Stark, L.~T. Muftuler, and C.~E. Stark, ``Distinct pattern separation related transfer functions in human ca3/dentate and ca1 revealed using high-resolution fmri and variable mnemonic similarity,'' {\em Learning \& memory}, vol.~18, no.~1, pp.~15--18, 2011.

\bibitem{Rust12978}
N.~C. Rust and J.~J. DiCarlo, ``Selectivity and tolerance ({\textquotedblleft}invariance{\textquotedblright}) both increase as visual information propagates from cortical area v4 to it,'' {\em Journal of Neuroscience}, vol.~30, no.~39, pp.~12978--12995, 2010.

\bibitem{dicarlo_how_2012}
J.~DiCarlo, D.~Zoccolan, and N.~Rust, ``How {Does} the {Brain} {Solve} {Visual} {Object} {Recognition}?,'' {\em Neuron}, vol.~73, pp.~415--434, Feb. 2012.
\newblock Publisher: Elsevier.

\bibitem{kravitz2013ventral}
D.~J. Kravitz, K.~S. Saleem, C.~I. Baker, L.~G. Ungerleider, and M.~Mishkin, ``The ventral visual pathway: an expanded neural framework for the processing of object quality,'' {\em Trends in cognitive sciences}, vol.~17, no.~1, pp.~26--49, 2013.

\bibitem{CHUNG2021137}
S.~Chung and L.~Abbott, ``Neural population geometry: An approach for understanding biological and artificial neural networks,'' {\em Current Opinion in Neurobiology}, vol.~70, pp.~137--144, 2021.
\newblock Computational Neuroscience.

\bibitem{wood2004stable}
S.~N. Wood, ``Stable and efficient multiple smoothing parameter estimation for generalized additive models,'' {\em Journal of the American Statistical Association}, vol.~99, no.~467, pp.~673--686, 2004.

\bibitem{park2009adaptive}
D.~C. Park and P.~Reuter-Lorenz, ``The adaptive brain: aging and neurocognitive scaffolding,'' {\em Annual review of psychology}, vol.~60, pp.~173--196, 2009.

\bibitem{naspi2023effects}
L.~Naspi, C.~Stensholt, A.~E. Karlsson, Z.~A. Monge, and R.~Cabeza, ``Effects of aging on successful object encoding: Enhanced semantic representations compensate for impaired visual representations,'' {\em Journal of Neuroscience}, vol.~43, no.~44, pp.~7337--7350, 2023.

\bibitem{chanales2021adaptive}
A.~J. Chanales, A.~G. Tremblay-McGaw, M.~L. Drascher, and B.~A. Kuhl, ``Adaptive repulsion of long-term memory representations is triggered by event similarity,'' {\em Psychological science}, vol.~32, no.~5, pp.~705--720, 2021.

\bibitem{prince2022improving}
J.~S. Prince, I.~Charest, J.~W. Kurzawski, J.~A. Pyles, M.~J. Tarr, and K.~N. Kay, ``Improving the accuracy of single-trial fmri response estimates using glmsingle,'' {\em Elife}, vol.~11, p.~e77599, 2022.

\bibitem{ngo2018ontogeny}
C.~T. Ngo, N.~S. Newcombe, and I.~R. Olson, ``The ontogeny of relational memory and pattern separation,'' {\em Developmental science}, vol.~21, no.~2, p.~e12556, 2018.

\bibitem{ngo2019building}
C.~T. Ngo, Y.~Lin, N.~S. Newcombe, and I.~R. Olson, ``Building up and wearing down episodic memory: Mnemonic discrimination and relational binding.,'' {\em Journal of Experimental Psychology: General}, vol.~148, no.~9, p.~1463, 2019.

\bibitem{canada2019s}
K.~L. Canada, C.~T. Ngo, N.~S. Newcombe, F.~Geng, and T.~Riggins, ``It’s all in the details: relations between young children’s developing pattern separation abilities and hippocampal subfield volumes,'' {\em Cerebral Cortex}, vol.~29, no.~8, pp.~3427--3433, 2019.

\bibitem{korkki2020healthy}
S.~M. Korkki, F.~R. Richter, P.~Jeyarathnarajah, and J.~S. Simons, ``Healthy ageing reduces the precision of episodic memory retrieval.,'' {\em Psychology and Aging}, vol.~35, no.~1, p.~124, 2020.

\bibitem{rosch1975cognitive}
E.~Rosch, ``Cognitive representations of semantic categories.,'' {\em Journal of experimental psychology: General}, vol.~104, no.~3, p.~192, 1975.

\bibitem{luo2007age}
L.~Luo, T.~Hendriks, and F.~I. Craik, ``Age differences in recollection: three patterns of enhanced encoding.,'' {\em Psychology and aging}, vol.~22, no.~2, p.~269, 2007.

\bibitem{dubova2021influences}
M.~Dubova and R.~L. Goldstone, ``The influences of category learning on perceptual reconstructions,'' {\em Cognitive Science}, vol.~45, no.~5, p.~e12981, 2021.

\bibitem{kramer2023features}
M.~A. Kramer, M.~N. Hebart, C.~I. Baker, and W.~A. Bainbridge, ``The features underlying the memorability of objects,'' {\em Science advances}, vol.~9, no.~17, p.~eadd2981, 2023.

\bibitem{cohen2025pattern}
S.~S. Cohen, C.~T. Ngo, I.~R. Olson, and N.~S. Newcombe, ``Pattern separation and pattern completion in early childhood,'' {\em Proceedings of the National Academy of Sciences}, vol.~122, no.~11, p.~e2416985122, 2025.

\bibitem{nemeczdifferent}
Z.~Nemecz, A.~Ily{\'e}s, L.~Kerekes, H.~Kis, M.~Werkle-Bergner, and A.~Keresztes, ``Different object features shape mnemonic discrimination in younger and older adults,''

\bibitem{ganguli2012compressed}
S.~Ganguli and H.~Sompolinsky, ``Compressed sensing, sparsity, and dimensionality in neuronal information processing and data analysis,'' {\em Annual review of neuroscience}, vol.~35, no.~1, pp.~485--508, 2012.

\bibitem{saxe2019information}
A.~M. Saxe, Y.~Bansal, J.~Dapello, M.~Advani, A.~Kolchinsky, B.~D. Tracey, and D.~D. Cox, ``On the information bottleneck theory of deep learning,'' {\em Journal of Statistical Mechanics: Theory and Experiment}, vol.~2019, no.~12, p.~124020, 2019.

\bibitem{schacter1999seven}
D.~L. Schacter, ``The seven sins of memory: insights from psychology and cognitive neuroscience.,'' {\em American psychologist}, vol.~54, no.~3, p.~182, 1999.

\bibitem{wimber2015retrieval}
M.~Wimber, A.~Alink, I.~Charest, N.~Kriegeskorte, and M.~C. Anderson, ``Retrieval induces adaptive forgetting of competing memories via cortical pattern suppression,'' {\em Nature neuroscience}, vol.~18, no.~4, pp.~582--589, 2015.

\bibitem{wang2019more}
T.~H. Wang, K.~Placek, and J.~A. Lewis-Peacock, ``More is less: increased processing of unwanted memories facilitates forgetting,'' {\em Journal of Neuroscience}, vol.~39, no.~18, pp.~3551--3560, 2019.

\bibitem{lynn2020abstract}
C.~W. Lynn, A.~E. Kahn, N.~Nyema, and D.~S. Bassett, ``Abstract representations of events arise from mental errors in learning and memory,'' {\em Nature communications}, vol.~11, no.~1, p.~2313, 2020.

\bibitem{lin2024images}
Q.~Lin, Z.~Li, J.~Lafferty, and I.~Yildirim, ``Images with harder-to-reconstruct visual representations leave stronger memory traces,'' {\em Nature human behaviour}, vol.~8, no.~7, pp.~1309--1320, 2024.

\bibitem{schacter20241373}
D.~L. Schacter, A.~C. Carpenter, A.~L. Devitt, and P.~P. Thakral, ``1373 memory errors and distortion,'' {\em The Oxford Handbook of Human Memory, Two Volume Pack: Foundations and Applications}, pp.~1373--1399, 2024.

\bibitem{wanjia2021abrupt}
G.~Wanjia, S.~E. Favila, G.~Kim, R.~J. Molitor, and B.~A. Kuhl, ``Abrupt hippocampal remapping signals resolution of memory interference,'' {\em Nature communications}, vol.~12, no.~1, p.~4816, 2021.

\bibitem{spens2024generative}
E.~Spens and N.~Burgess, ``A generative model of memory construction and consolidation,'' {\em Nature human behaviour}, vol.~8, no.~3, pp.~526--543, 2024.

\bibitem{reagh2014object}
Z.~M. Reagh and M.~A. Yassa, ``Object and spatial mnemonic interference differentially engage lateral and medial entorhinal cortex in humans,'' {\em Proceedings of the National Academy of Sciences}, vol.~111, no.~40, pp.~E4264--E4273, 2014.

\bibitem{pidgeon2016cortical}
L.~M. Pidgeon and A.~M. Morcom, ``Cortical pattern separation and item-specific memory encoding,'' {\em Neuropsychologia}, vol.~85, pp.~256--271, 2016.

\bibitem{wais2017retrieval}
P.~E. Wais, S.~Jahanikia, D.~Steiner, C.~E. Stark, and A.~Gazzaley, ``Retrieval of high-fidelity memory arises from distributed cortical networks,'' {\em NeuroImage}, vol.~149, pp.~178--189, 2017.

\bibitem{reagh2018functional}
Z.~M. Reagh, J.~A. Noche, N.~J. Tustison, D.~Delisle, E.~A. Murray, and M.~A. Yassa, ``Functional imbalance of anterolateral entorhinal cortex and hippocampal dentate/ca3 underlies age-related object pattern separation deficits,'' {\em Neuron}, vol.~97, no.~5, pp.~1187--1198, 2018.

\bibitem{zhao2021adaptive}
Y.~Zhao, A.~J. Chanales, and B.~A. Kuhl, ``Adaptive memory distortions are predicted by feature representations in parietal cortex,'' {\em Journal of Neuroscience}, vol.~41, no.~13, pp.~3014--3024, 2021.

\bibitem{johnson2021rodent}
S.~A. Johnson, S.~Zequeira, S.~M. Turner, A.~P. Maurer, J.~L. Bizon, and S.~N. Burke, ``Rodent mnemonic similarity task performance requires the prefrontal cortex,'' {\em Hippocampus}, vol.~31, no.~7, pp.~701--716, 2021.

\bibitem{amer2023extra}
T.~Amer and L.~Davachi, ``Extra-hippocampal contributions to pattern separation,'' {\em elife}, vol.~12, p.~e82250, 2023.

\bibitem{wanjia2025repulsion}
G.~Wanjia, S.~Han, and B.~A. Kuhl, ``Repulsion of hippocampal representations driven by distinct internal beliefs,'' {\em Current Biology}, 2025.

\bibitem{barlow1961possible}
H.~B. Barlow {\em et~al.}, ``Possible principles underlying the transformation of sensory messages,'' {\em Sensory communication}, vol.~1, no.~01, pp.~217--233, 1961.

\bibitem{attneave1954some}
F.~Attneave, ``Some informational aspects of visual perception.,'' {\em Psychological review}, vol.~61, no.~3, p.~183, 1954.

\bibitem{li2014understanding}
Z.~Li, {\em Understanding vision: theory, models, and data}.
\newblock Oxford University Press (UK), 2014.

\bibitem{koch2006much}
K.~Koch, J.~McLean, R.~Segev, M.~A. Freed, M.~J. Berry, V.~Balasubramanian, and P.~Sterling, ``How much the eye tells the brain,'' {\em Current biology}, vol.~16, no.~14, pp.~1428--1434, 2006.

\bibitem{pitkow2012decorrelation}
X.~Pitkow and M.~Meister, ``Decorrelation and efficient coding by retinal ganglion cells,'' {\em Nature neuroscience}, vol.~15, no.~4, pp.~628--635, 2012.

\bibitem{harris2015energy}
J.~J. Harris, R.~Jolivet, E.~Engl, and D.~Attwell, ``Energy-efficient information transfer by visual pathway synapses,'' {\em Current Biology}, vol.~25, no.~24, pp.~3151--3160, 2015.

\bibitem{owen2024high}
L.~L. Owen and J.~R. Manning, ``High-level cognition is supported by information-rich but compressible brain activity patterns,'' {\em Proceedings of the National Academy of Sciences}, vol.~121, no.~35, p.~e2400082121, 2024.

\bibitem{tang2019effective}
E.~Tang, M.~G. Mattar, C.~Giusti, D.~M. Lydon-Staley, S.~L. Thompson-Schill, and D.~S. Bassett, ``Effective learning is accompanied by high-dimensional and efficient representations of neural activity,'' {\em Nature neuroscience}, vol.~22, no.~6, pp.~1000--1009, 2019.

\bibitem{kolchinsky2019nonlinear}
A.~Kolchinsky, B.~D. Tracey, and D.~H. Wolpert, ``Nonlinear information bottleneck,'' {\em Entropy}, vol.~21, no.~12, p.~1181, 2019.

\bibitem{dubova2023excess}
M.~Dubova and S.~J. Sloman, ``Excess capacity learning,'' in {\em Proceedings of the annual meeting of the cognitive science society}, vol.~45, 2023.

\bibitem{ito2023multitask}
T.~Ito and J.~D. Murray, ``Multitask representations in the human cortex transform along a sensory-to-motor hierarchy,'' {\em Nature Neuroscience}, vol.~26, no.~2, pp.~306--315, 2023.

\bibitem{nigam2024predictions}
T.~Nigam and C.~M. Schwiedrzik, ``Predictions enable top-down pattern separation in the macaque face-processing hierarchy,'' {\em Nature Communications}, vol.~15, no.~1, p.~7196, 2024.

\bibitem{whittington2022disentanglement}
J.~C. Whittington, W.~Dorrell, S.~Ganguli, and T.~E. Behrens, ``Disentanglement with biological constraints: A theory of functional cell types,'' {\em arXiv preprint arXiv:2210.01768}, 2022.

\bibitem{kerren2025exploring}
C.~Kerr{\'e}n, D.~Reznik, C.~F. Doeller, and B.~J. Griffiths, ``Exploring the role of dimensionality transformation in episodic memory,'' {\em Trends in Cognitive Sciences}, vol.~29, no.~7, pp.~614--626, 2025.

\bibitem{ranganath2003neural}
C.~Ranganath and G.~Rainer, ``Neural mechanisms for detecting and remembering novel events,'' {\em Nature Reviews Neuroscience}, vol.~4, no.~3, pp.~193--202, 2003.

\bibitem{jaegle2019visual}
A.~Jaegle, V.~Mehrpour, and N.~Rust, ``Visual novelty, curiosity, and intrinsic reward in machine learning and the brain,'' {\em Current opinion in neurobiology}, vol.~58, pp.~167--174, 2019.

\bibitem{lisman2005hippocampal}
J.~E. Lisman and A.~A. Grace, ``The hippocampal-vta loop: controlling the entry of information into long-term memory,'' {\em Neuron}, vol.~46, no.~5, pp.~703--713, 2005.

\bibitem{hasselmo1997free}
M.~E. Hasselmo and B.~P. Wyble, ``Free recall and recognition in a network model of the hippocampus: simulating effects of scopolamine on human memory function,'' {\em Behavioural brain research}, vol.~89, no.~1-2, pp.~1--34, 1997.

\bibitem{zbontar2021barlow}
J.~Zbontar, L.~Jing, I.~Misra, Y.~LeCun, and S.~Deny, ``Barlow twins: Self-supervised learning via redundancy reduction,'' in {\em International conference on machine learning}, pp.~12310--12320, PMLR, 2021.

\bibitem{carr2013top}
V.~A. Carr, S.~A. Engel, and B.~J. Knowlton, ``Top-down modulation of hippocampal encoding activity as measured by high-resolution functional mri,'' {\em Neuropsychologia}, vol.~51, no.~10, pp.~1829--1837, 2013.

\bibitem{aly2016attention}
M.~Aly and N.~B. Turk-Browne, ``Attention promotes episodic encoding by stabilizing hippocampal representations,'' {\em Proceedings of the National Academy of Sciences}, vol.~113, no.~4, pp.~E420--E429, 2016.

\bibitem{son2024brief}
G.~Son, D.~B. Walther, and M.~L. Mack, ``Brief category learning distorts perceptual space for complex scenes,'' {\em Psychonomic Bulletin \& Review}, vol.~31, no.~5, pp.~2234--2248, 2024.

\bibitem{mack2024discrimination}
M.~L. Mack and T.~J. Palmeri, ``Discrimination, recognition, and classification,'' 2024.

\bibitem{bar2006top}
M.~Bar, K.~S. Kassam, A.~S. Ghuman, J.~Boshyan, A.~M. Schmid, A.~M. Dale, M.~S. H{\"a}m{\"a}l{\"a}inen, K.~Marinkovic, D.~L. Schacter, B.~R. Rosen, {\em et~al.}, ``Top-down facilitation of visual recognition,'' {\em Proceedings of the national academy of sciences}, vol.~103, no.~2, pp.~449--454, 2006.

\bibitem{hindy2016linking}
N.~C. Hindy, F.~Y. Ng, and N.~B. Turk-Browne, ``Linking pattern completion in the hippocampus to predictive coding in visual cortex,'' {\em Nature neuroscience}, vol.~19, no.~5, pp.~665--667, 2016.

\bibitem{kim2020functions}
J.~G. Kim, E.~Gregory, B.~Landau, M.~McCloskey, N.~B. Turk-Browne, and S.~Kastner, ``Functions of ventral visual cortex after bilateral medial temporal lobe damage,'' {\em Progress in neurobiology}, vol.~191, p.~101819, 2020.

\bibitem{isola2011understanding}
P.~Isola, D.~Parikh, A.~Torralba, and A.~Oliva, ``Understanding the intrinsic memorability of images,'' {\em Advances in neural information processing systems}, vol.~24, 2011.

\bibitem{khosla2012memorability}
A.~Khosla, J.~Xiao, A.~Torralba, and A.~Oliva, ``Memorability of image regions,'' {\em Advances in neural information processing systems}, vol.~25, 2012.

\bibitem{rust2020understanding}
N.~C. Rust and V.~Mehrpour, ``Understanding image memorability,'' {\em Trends in cognitive sciences}, vol.~24, no.~7, pp.~557--568, 2020.

\bibitem{bylinskii2021memorability}
Z.~Bylinskii, L.~Goetschalckx, A.~Newman, and A.~Oliva, ``Memorability: An image-computable measure of information utility,'' in {\em Human perception of visual information: Psychological and computational perspectives}, pp.~207--239, Springer, 2021.

\bibitem{revsine2025memorability}
C.~Revsine and W.~A. Bainbridge, ``Memorability reflects statistical regularities of the environment,'' {\em Current Opinion in Neurobiology}, vol.~94, p.~103095, 2025.

\bibitem{murphy2022interpretability}
K.~A. Murphy and D.~S. Bassett, ``Interpretability with full complexity by constraining feature information,'' {\em arXiv preprint arXiv:2211.17264}, 2022.

\bibitem{fang2025humans}
Z.~Fang and C.~R. Sims, ``Humans learn generalizable representations through efficient coding,'' {\em Nature Communications}, vol.~16, no.~1, p.~3989, 2025.

\bibitem{kahn2004functional}
I.~Kahn, L.~Davachi, and A.~D. Wagner, ``Functional-neuroanatomic correlates of recollection: implications for models of recognition memory,'' {\em Journal of Neuroscience}, vol.~24, no.~17, pp.~4172--4180, 2004.

\bibitem{kuhl2013dissociable}
B.~A. Kuhl, M.~K. Johnson, and M.~M. Chun, ``Dissociable neural mechanisms for goal-directed versus incidental memory reactivation,'' {\em Journal of Neuroscience}, vol.~33, no.~41, pp.~16099--16109, 2013.

\bibitem{richter2016distinct}
F.~R. Richter, R.~A. Cooper, P.~M. Bays, and J.~S. Simons, ``Distinct neural mechanisms underlie the success, precision, and vividness of episodic memory,'' {\em elife}, vol.~5, p.~e18260, 2016.

\bibitem{favila2018parietal}
S.~E. Favila, R.~Samide, S.~C. Sweigart, and B.~A. Kuhl, ``Parietal representations of stimulus features are amplified during memory retrieval and flexibly aligned with top-down goals,'' {\em Journal of Neuroscience}, vol.~38, no.~36, pp.~7809--7821, 2018.

\bibitem{yang2024pattern}
Z.~Yang, X.~Zhuang, K.~A. Koenig, J.~B. Leverenz, T.~Curran, M.~J. Lowe, and D.~Cordes, ``Pattern separation involves regions beyond the hippocampus in non-demented elderly individuals: A 7t object lure task fmri study,'' {\em Imaging Neuroscience}, vol.~2, pp.~1--15, 2024.

\bibitem{nosofsky2000exemplar}
R.~M. Nosofsky and M.~K. Johansen, ``Exemplar-based accounts of “multiple-system” phenomena in perceptual categorization,'' {\em Psychonomic Bulletin \& Review}, vol.~7, no.~3, pp.~375--402, 2000.

\bibitem{sigala2002visual}
N.~Sigala and N.~K. Logothetis, ``Visual categorization shapes feature selectivity in the primate temporal cortex,'' {\em Nature}, vol.~415, no.~6869, pp.~318--320, 2002.

\bibitem{folstein2013category}
J.~R. Folstein, T.~J. Palmeri, and I.~Gauthier, ``Category learning increases discriminability of relevant object dimensions in visual cortex,'' {\em Cerebral Cortex}, vol.~23, no.~4, pp.~814--823, 2013.

\bibitem{folstein2015category}
J.~R. Folstein, T.~J. Palmeri, A.~E. Van~Gulick, and I.~Gauthier, ``Category learning stretches neural representations in visual cortex,'' {\em Current directions in psychological science}, vol.~24, no.~1, pp.~17--23, 2015.

\bibitem{klippenstein2020neural}
J.~L. Klippenstein, S.~M. Stark, C.~E. Stark, and I.~J. Bennett, ``Neural substrates of mnemonic discrimination: A whole-brain fmri investigation,'' {\em Brain and Behavior}, vol.~10, no.~3, p.~e01560, 2020.

\bibitem{norman2003modeling}
K.~A. Norman and R.~C. O'Reilly, ``Modeling hippocampal and neocortical contributions to recognition memory: a complementary-learning-systems approach.,'' {\em Psychological review}, vol.~110, no.~4, p.~611, 2003.

\bibitem{rotshtein2005morphing}
P.~Rotshtein, R.~N. Henson, A.~Treves, J.~Driver, and R.~J. Dolan, ``Morphing marilyn into maggie dissociates physical and identity face representations in the brain,'' {\em Nature neuroscience}, vol.~8, no.~1, pp.~107--113, 2005.

\bibitem{paleja2014two}
M.~Paleja, T.~A. Girard, K.~A. Herdman, and B.~K. Christensen, ``Two distinct neural networks functionally connected to the human hippocampus during pattern separation tasks,'' {\em Brain and Cognition}, vol.~92, pp.~101--111, 2014.

\bibitem{gattas2023theta}
S.~Gattas, M.~S. Larson, L.~Mnatsakanyan, I.~Sen-Gupta, S.~Vadera, A.~L. Swindlehurst, P.~E. Rapp, J.~J. Lin, and M.~A. Yassa, ``Theta mediated dynamics of human hippocampal-neocortical learning systems in memory formation and retrieval,'' {\em Nature communications}, vol.~14, no.~1, p.~8505, 2023.

\bibitem{davidson2019older}
P.~S. Davidson, P.~Vidjen, S.~Trincao-Batra, and C.~A. Collin, ``Older adults’ lure discrimination difficulties on the mnemonic similarity task are significantly correlated with their visual perception,'' {\em The Journals of Gerontology: Series B}, vol.~74, no.~8, pp.~1298--1307, 2019.

\bibitem{barrow1978computer}
H.~Barrow, J.~Tenenbaum, A.~Hanson, and E.~Riseman, {\em Computer vision systems}.
\newblock 1978.

\bibitem{yuille2006vision}
A.~Yuille and D.~Kersten, ``Vision as bayesian inference: analysis by synthesis?,'' {\em Trends in cognitive sciences}, vol.~10, no.~7, pp.~301--308, 2006.

\bibitem{olshausen2014perception}
B.~A. Olshausen, G.~Mangun, and M.~Gazzaniga, ``Perception as an inference problem,'' {\em The cognitive neurosciences}, pp.~295--304, 2014.

\bibitem{harlow2016distinguishing}
I.~M. Harlow and A.~P. Yonelinas, ``Distinguishing between the success and precision of recollection,'' {\em Memory}, vol.~24, no.~1, pp.~114--127, 2016.

\bibitem{jang2024improved}
H.~Jang and F.~Tong, ``Improved modeling of human vision by incorporating robustness to blur in convolutional neural networks,'' {\em Nature Communications}, vol.~15, no.~1, p.~1989, 2024.

\bibitem{rotello2000associative}
C.~M. Rotello and E.~Heit, ``Associative recognition: A case of recall-to-reject processing,'' {\em Memory \& Cognition}, vol.~28, no.~6, pp.~907--922, 2000.

\bibitem{gallo2006two}
D.~A. Gallo, D.~M. Bell, J.~S. Beier, and D.~L. Schacter, ``Two types of recollection-based monitoring in younger and older adults: Recall-to-reject and the distinctiveness heuristic,'' {\em Memory}, vol.~14, no.~6, pp.~730--741, 2006.

\bibitem{wahlheimrole}
C.~Wahlheim and L.~Richmond, ``A role for pattern completion in lure rejection evinced in subsequent order memory,''

\bibitem{wahlheimmnemonic}
C.~Wahlheim, I.~Dobbins, and B.~Wellons, ``Mnemonic discrimination language evinces recollection rejection of similar lures,''

\bibitem{DiRisio2025.09.22.677855}
G.~F. DiRisio, C.~Xue, and M.~R. Cohen, ``Neuronal signatures of successful one-shot memory in mid-level visual cortex,'' {\em bioRxiv}, 2025.

\bibitem{wheeler2000memory}
M.~E. Wheeler, S.~E. Petersen, and R.~L. Buckner, ``Memory's echo: vivid remembering reactivates sensory-specific cortex,'' {\em Proceedings of the National Academy of Sciences}, vol.~97, no.~20, pp.~11125--11129, 2000.

\bibitem{kuhl2011fidelity}
B.~A. Kuhl, J.~Rissman, M.~M. Chun, and A.~D. Wagner, ``Fidelity of neural reactivation reveals competition between memories,'' {\em Proceedings of the National Academy of Sciences}, vol.~108, no.~14, pp.~5903--5908, 2011.

\bibitem{mcnaughton1995there}
B.~McNaughton and R.~O’reilly, ``Why there are complementary learning systems in the hippocampus and neocortex: insights from the successes and failures of connectionist models of learning and memory,'' {\em Psychol Rev}, 1995.

\bibitem{bein2025schemas}
O.~Bein and Y.~Niv, ``Schemas, reinforcement learning and the medial prefrontal cortex,'' {\em Nature Reviews Neuroscience}, vol.~26, no.~3, pp.~141--157, 2025.

\bibitem{giguere2013limits}
G.~Gigu{\`e}re and B.~C. Love, ``Limits in decision making arise from limits in memory retrieval,'' {\em Proceedings of the National Academy of Sciences}, vol.~110, no.~19, pp.~7613--7618, 2013.

\bibitem{wolfe2021guided}
J.~M. Wolfe, ``Guided search 6.0: An updated model of visual search,'' {\em Psychonomic bulletin \& review}, vol.~28, no.~4, pp.~1060--1092, 2021.

\bibitem{brady2009compression}
T.~F. Brady, T.~Konkle, and G.~A. Alvarez, ``Compression in visual working memory: using statistical regularities to form more efficient memory representations.,'' {\em Journal of Experimental Psychology: General}, vol.~138, no.~4, p.~487, 2009.

\bibitem{sims2015cost}
C.~R. Sims, ``The cost of misremembering: Inferring the loss function in visual working memory,'' {\em Journal of vision}, vol.~15, no.~3, pp.~2--2, 2015.

\bibitem{simoncelli2001natural}
E.~P. Simoncelli and B.~A. Olshausen, ``Natural image statistics and neural representation,'' {\em Annual review of neuroscience}, vol.~24, no.~1, pp.~1193--1216, 2001.

\bibitem{karklin2011efficient}
Y.~Karklin and E.~Simoncelli, ``Efficient coding of natural images with a population of noisy linear-nonlinear neurons,'' {\em Advances in neural information processing systems}, vol.~24, 2011.

\bibitem{mukherjee2023evaluating}
K.~Mukherjee, X.~Lu, H.~Huey, Y.~Vinker, R.~Aguina-Kang, A.~Shamir, and J.~E. Fan, ``Evaluating machine comprehension of sketch meaning at different levels of abstraction,'' in {\em Proceedings of the Annual Meeting of the Cognitive Science Society}, vol.~45, 2023.

\bibitem{long2024parallel}
B.~Long, J.~E. Fan, H.~Huey, Z.~Chai, and M.~C. Frank, ``Parallel developmental changes in children’s production and recognition of line drawings of visual concepts,'' {\em Nature Communications}, vol.~15, no.~1, p.~1191, 2024.

\bibitem{recanatesi2022scale}
S.~Recanatesi, S.~Bradde, V.~Balasubramanian, N.~A. Steinmetz, and E.~Shea-Brown, ``A scale-dependent measure of system dimensionality,'' {\em Patterns}, vol.~3, no.~8, 2022.

\bibitem{hebart2020revealing}
M.~N. Hebart, C.~Y. Zheng, F.~Pereira, and C.~I. Baker, ``Revealing the multidimensional mental representations of natural objects underlying human similarity judgements,'' {\em Nature human behaviour}, vol.~4, no.~11, pp.~1173--1185, 2020.

\bibitem{fu2023dreamsim}
S.~Fu, N.~Tamir, S.~Sundaram, L.~Chai, R.~Zhang, T.~Dekel, and P.~Isola, ``Dreamsim: Learning new dimensions of human visual similarity using synthetic data,'' {\em arXiv preprint arXiv:2306.09344}, 2023.

\bibitem{kim2014pruning}
G.~Kim, J.~A. Lewis-Peacock, K.~A. Norman, and N.~B. Turk-Browne, ``Pruning of memories by context-based prediction error,'' {\em Proceedings of the National Academy of Sciences}, vol.~111, no.~24, pp.~8997--9002, 2014.

\bibitem{allen2022massive}
E.~J. Allen, G.~St-Yves, Y.~Wu, J.~L. Breedlove, J.~S. Prince, L.~T. Dowdle, M.~Nau, B.~Caron, F.~Pestilli, I.~Charest, {\em et~al.}, ``A massive 7t fmri dataset to bridge cognitive neuroscience and artificial intelligence,'' {\em Nature neuroscience}, vol.~25, no.~1, pp.~116--126, 2022.

\bibitem{murphy2024information}
K.~A. Murphy and D.~S. Bassett, ``Information decomposition in complex systems via machine learning,'' {\em Proceedings of the National Academy of Sciences}, vol.~121, no.~13, p.~e2312988121, 2024.

\bibitem{kim2013assessing}
J.~Kim and M.~A. Yassa, ``Assessing recollection and familiarity of similar lures in a behavioral pattern separation task,'' {\em Hippocampus}, vol.~23, no.~4, pp.~287--294, 2013.

\bibitem{kirwan2007overcoming}
C.~B. Kirwan and C.~E. Stark, ``Overcoming interference: An fmri investigation of pattern separation in the medial temporal lobe,'' {\em Learning \& Memory}, vol.~14, no.~9, pp.~625--633, 2007.

\bibitem{stark2021playing}
C.~E. Stark, G.~D. Clemenson, U.~Aluru, N.~Hatamian, and S.~M. Stark, ``Playing minecraft improves hippocampal-associated memory for details in middle aged adults,'' {\em Frontiers in sports and active living}, vol.~3, p.~685286, 2021.

\bibitem{kingma2013auto}
D.~P. Kingma, M.~Welling, {\em et~al.}, ``Auto-encoding variational bayes,'' 2013.

\bibitem{xu2013evaluation}
J.~Xu, S.~Moeller, E.~J. Auerbach, J.~Strupp, S.~M. Smith, D.~A. Feinberg, E.~Yacoub, and K.~U{\u{g}}urbil, ``Evaluation of slice accelerations using multiband echo planar imaging at 3 t,'' {\em Neuroimage}, vol.~83, pp.~991--1001, 2013.

\bibitem{cox1996afni}
R.~W. Cox, ``Afni: software for analysis and visualization of functional magnetic resonance neuroimages,'' {\em Computers and Biomedical research}, vol.~29, no.~3, pp.~162--173, 1996.

\bibitem{rokem2020fractional}
A.~Rokem and K.~Kay, ``Fractional ridge regression: a fast, interpretable reparameterization of ridge regression,'' {\em GigaScience}, vol.~9, no.~12, p.~giaa133, 2020.

\bibitem{scikitlearn}
F.~Pedregosa, G.~Varoquaux, A.~Gramfort, V.~Michel, B.~Thirion, O.~Grisel, M.~Blondel, P.~Prettenhofer, R.~Weiss, V.~Dubourg, J.~Vanderplas, A.~Passos, D.~Cournapeau, M.~Brucher, M.~Perrot, and E.~Duchesnay, ``Scikit-learn: Machine learning in {P}ython,'' {\em Journal of Machine Learning Research}, vol.~12, pp.~2825--2830, 2011.

\bibitem{kraskov2004estimating}
A.~Kraskov, H.~St{\"o}gbauer, and P.~Grassberger, ``Estimating mutual information,'' {\em Physical Review E—Statistical, Nonlinear, and Soft Matter Physics}, vol.~69, no.~6, p.~066138, 2004.

\bibitem{lizier2014jidt}
J.~T. Lizier, ``Jidt: An information-theoretic toolkit for studying the dynamics of complex systems,'' {\em Frontiers in Robotics and AI}, vol.~1, p.~11, 2014.

\bibitem{mitchell2013gendered}
S.~M. Mitchell, S.~Lange, and H.~Brus, ``Gendered citation patterns in international relations journals,'' {\em International Studies Perspectives}, vol.~14, no.~4, pp.~485--492, 2013.

\bibitem{dion2018gendered}
M.~L. Dion, J.~L. Sumner, and S.~M. Mitchell, ``Gendered citation patterns across political science and social science methodology fields,'' {\em Political Analysis}, vol.~26, no.~3, pp.~312--327, 2018.

\bibitem{caplar2017quantitative}
N.~Caplar, S.~Tacchella, and S.~Birrer, ``Quantitative evaluation of gender bias in astronomical publications from citation counts,'' {\em Nature Astronomy}, vol.~1, no.~6, p.~0141, 2017.

\bibitem{maliniak2013gender}
D.~Maliniak, R.~Powers, and B.~F. Walter, ``The gender citation gap in international relations,'' {\em International Organization}, vol.~67, no.~4, pp.~889--922, 2013.

\bibitem{Dworkin2020.01.03.894378}
J.~D. Dworkin, K.~A. Linn, E.~G. Teich, P.~Zurn, R.~T. Shinohara, and D.~S. Bassett, ``The extent and drivers of gender imbalance in neuroscience reference lists,'' {\em bioRxiv}, 2020.

\bibitem{bertolero2021racial}
M.~A. Bertolero, J.~D. Dworkin, S.~U. David, C.~L. Lloreda, P.~Srivastava, J.~Stiso, D.~Zhou, K.~Dzirasa, D.~A. Fair, A.~N. Kaczkurkin, B.~J. Marlin, D.~Shohamy, L.~Q. Uddin, P.~Zurn, and D.~S. Bassett, ``Racial and ethnic imbalance in neuroscience reference lists and intersections with gender,'' {\em bioRxiv}, 2020.

\bibitem{wang2021gendered}
X.~Wang, J.~D. Dworkin, D.~Zhou, J.~Stiso, E.~B. Falk, D.~S. Bassett, P.~Zurn, and D.~M. Lydon-Staley, ``Gendered citation practices in the field of communication,'' {\em Annals of the International Communication Association}, 2021.

\bibitem{chatterjee2021gender}
P.~Chatterjee and R.~M. Werner, ``Gender disparity in citations in high-impact journal articles,'' {\em JAMA Netw Open}, vol.~4, no.~7, p.~e2114509, 2021.

\bibitem{fulvio2021imbalance}
J.~M. Fulvio, I.~Akinnola, and B.~R. Postle, ``Gender (im)balance in citation practices in cognitive neuroscience,'' {\em J Cogn Neurosci}, vol.~33, no.~1, pp.~3--7, 2021.

\bibitem{zhou_dale_2020_3672110}
D.~Zhou, E.~J. Cornblath, J.~Stiso, E.~G. Teich, J.~D. Dworkin, A.~S. Blevins, and D.~S. Bassett, ``Gender diversity statement and code notebook v1.0,'' Feb. 2020.

\bibitem{ambekar2009name}
A.~Ambekar, C.~Ward, J.~Mohammed, S.~Male, and S.~Skiena, ``Name-ethnicity classification from open sources,'' in {\em Proceedings of the 15th ACM SIGKDD international conference on Knowledge Discovery and Data Mining}, pp.~49--58, 2009.

\bibitem{sood2018predicting}
G.~Sood and S.~Laohaprapanon, ``Predicting race and ethnicity from the sequence of characters in a name,'' {\em arXiv preprint arXiv:1805.02109}, 2018.

\end{thebibliography}

\end{document}